\pdfoutput=1
\documentclass[11pt,a4paper]{article}

\usepackage{jheppub}
\usepackage[mathscr]{euscript}

\usepackage{mathtools,amssymb,latexsym,bm,amsfonts}

\hypersetup{
	bookmarksnumbered=true,
	bookmarksopen=true,
	bookmarksopenlevel=2
}

\usepackage{graphicx,graphbox}
\usepackage{longtable}
\usepackage{subfig}
\usepackage{comment}
\usepackage{float}
\usepackage[normalem]{ulem}
\usepackage{array}
\usepackage{enumitem}
\usepackage{relsize}
\usepackage{graphicx}

\numberwithin{equation}{section}
\allowdisplaybreaks

\usepackage[defaultlines=2,all]{nowidow}
\makeatletter
\def\@fpheader{\phantom{Prepared for submission to JHEP}}
\makeatother

\newcommand{\bea}{{\begin{eqnarray}}}
\newcommand{\eea}{{\end{eqnarray}}}

\newcommand{\be}{\begin{equation}}
\newcommand{\ee}{\end{equation}}
\newcommand{\bpm}{\begin{pmatrix}}
\newcommand{\epm}{\end{pmatrix}}

\newcommand{\beqn}{\begin{eqnarray}}
\newcommand{\eeqn}{\end{eqnarray}}

\newcommand{\p}{\partial}

\newcommand{\ba}{\begin{aligned}}
\newcommand{\ea}{\end{aligned}}
\newcommand{\bi}{\begin{enumerate}}
\newcommand{\ei}{\end{enumerate}}

\newcommand{\eq}[1]{\begin{align}#1\end{align}}
\newcommand{\eqsp}[1]{\begin{equation}\begin{split}#1\end{split}\end{equation}}

\let\textcite\cite
\newcommand{\abs}[1]{\left|#1\right|}

\let\ave\aqty
\newcommand{\pqty}[1]{\left(#1\right)}

\newcommand{\var}[1]{\mathinner{\delta #1}}

\newcommand{\mcal}[1]{\mathcal{#1}}

\newcommand{\thinmspace}[1][.5]{\mspace{#1\thinmuskip}}

\newcommand{\pdd}[1]{\partial_{#1}}

\newcommand{\pd}{\partial}

\def\til#1{\tilde{#1}}

\def\p{\partial}

\def\a{\alpha}

\def\d{\delta}

\def\vp{\varphi}

\title{Glue-on AdS holography for $T\bar T$-deformed CFTs}
\author{Luis Apolo$^{\,\alpha,\,\beta}$,}
\author{Peng-Xiang Hao$^{\,\gamma}$,}
\author{Wen-Xin Lai$^{\,\gamma,\,\tau}$,}
\author{Wei Song$^{\,\gamma,\,\tau}$}

\affiliation[\ensuremath{\alpha}]{Institute for Theoretical Physics, University of Amsterdam, 1090GL Amsterdam, The Netherlands}
\affiliation[\ensuremath{\beta}]{Beijing Institute of Mathematical Sciences and Applications, Beijing 101408, China}
\affiliation[\ensuremath{\gamma}]{Yau Mathematical Sciences Center, Tsinghua University, Beijing 100084, China}
\affiliation[\ensuremath{\tau}]{Department of Mathematical Sciences, Tsinghua University, Beijing 100084, China}

\emailAdd{l.a.apolo@uva.nl}
\emailAdd{pxhao@mail.tsinghua.edu.cn}
\emailAdd{laiwx19@mails.tsinghua.edu.cn}
\emailAdd{wsong2014@mail.tsinghua.edu.cn}

\abstract{%
The $T\bar T$ deformation is a solvable irrelevant deformation whose properties depend on the sign of the deformation parameter $\mu$. In particular, $T\bar T$-deformed CFTs with $\mu<0$ have been proposed to be holographically dual to Einstein gravity where the metric satisfies Dirichlet boundary conditions at a finite cutoff surface. In this paper, we put forward a holographic proposal for $T\bar T$-deformed CFTs with $\mu>0$, in which case the bulk geometry is constructed by gluing a patch of AdS$_3$ to the original spacetime. As evidence, we show that the $T\bar T$ trace flow equation, the spectrum on the cylinder, and the partition function on the torus and the sphere, among other results, can all be reproduced from bulk calculations in glue-on AdS$_3$.
}

\usepackage{xspace}
\newcommand{\TTbar}{\texorpdfstring{\ensuremath{T\bar{T}}}{TTbar}\xspace}

\begin{document}
\maketitle
\flushbottom

\section{Introduction}

The AdS/CFT correspondence \cite{Maldacena:1997re,Witten:1998qj,Gubser:1998bc} provides a holographic description of quantum gravity in asymptotically  anti de Sitter (AdS) spacetimes in terms of a conformal field theory (CFT) at the asymptotic boundary. While the study of AdS/CFT has been fruitful, e.g.~in the counting of black hole microstates \cite{Strominger:1996sh} and the computation of entanglement entropy \cite{Ryu:2006bv,Hubeny:2007xt}, it does not directly apply to other physically relevant backgrounds such as flat or de Sitter spacetimes, that are not asymptotically AdS.

In order to gain a better understanding of quantum gravity and holography, it is desirable to extend the AdS/CFT correspondence beyond its original domain of validity.
One way of accomplishing this is by deforming both the bulk and boundary sides of the correspondence. On the boundary side, relevant deformations of CFTs have been extensively studied in the literature, see e.g.~\cite{Leigh:1995ep,Girardello:1998pd,Freedman:1999gp}. In the bulk, these deformations modify the classical solution in the interior of the spacetime, but keep the asymptotic boundary intact. Deforming the asymptotic region of the spacetime, on the other hand, corresponds to introducing irrelevant deformations in the dual CFT, which are notoriously difficult to deal with except for a few exceptions.

The \TTbar deformation is an example of a tractable irrelevant deformation that induces a flow towards the UV and is defined by~\cite{Zamolodchikov:2004ce,Smirnov:2016lqw,Cavaglia:2016oda}
\eq{ \label{TTbardefintro}
\partial_\mu I &= {8 \pi} \int d^2x \, T\bar T =  {\pi} \int d^2x\big( T^{ij}T_{ij}- (T^i_i)^2 \big),
}
where $\mu$ is the deformation parameter and $T_{ij}$ is the instantaneous stress energy tensor of the deformed theory whose action is denoted by $I$. The $T\bar T$ operator is well defined in any translationally invariant two-dimensional quantum field theory (QFT).
The universality of the $T\bar T$ operator translates into universal features of the deformation that do not depend on the details of the undeformed theory. For example, the S-matrix of any \TTbar-deformed QFT can be written in terms of the S-matrix of the undeformed theory~\cite{Dubovsky:2012wk,Dubovsky:2013ira}. Similarly, the spectrum of \TTbar-deformed CFTs on the cylinder can be expressed in terms of the undeformed spectrum~\cite{Smirnov:2016lqw,Cavaglia:2016oda}. Other interesting features of the \TTbar deformation include the modular invariance of the partition function~\cite{Datta:2018thy,Aharony:2018bad}, connections to two-dimensional quantum gravity \cite{Dubovsky:2017cnj,Cardy:2018sdv}, reformulations as critical and non-critical string theory \cite{Callebaut:2019omt,Tolley:2019nmm}, supersymmetric generalizations~\cite{Baggio:2018rpv,Chang:2018dge,Jiang:2019hux,Chang:2019kiu}, as well as generalizations to higher \cite{Hartman:2018tkw,Taylor:2018xcy} and lower dimensions \cite{Gross:2019ach}, among others \cite{Dubovsky:2018bmo,Jafari:2019qns,Belin:2020oib}.

The universal nature of the $T\bar T$-deformation makes it readily applicable within the bottom-up approach to the AdS$_3$/CFT$_2$ correspondence. In this approach, the bulk is described by three-dimensional Einstein gravity with a negative cosmological constant, while the theory at the boundary is a two-dimensional CFT with a large central charge and sparse light spectrum, among other properties. The authors of \cite{McGough:2016lol} proposed that turning on the $T\bar T$ deformation with $\mu < 0$ is equivalent to placing the gravitational theory in a finite box with Dirichlet boundary conditions at a cutoff surface.\footnote{Subtleties arise when the bulk theory contains matter fields, in which case the cutoff AdS proposal requires additional double-trace deformations in the dual CFT \cite{Kraus:2018xrn,Hartman:2018tkw,Guica:2019nzm}. } It is important to note that the location of the cutoff surface is proportional to $-\mu$, so that the cutoff AdS proposal works only for negative values of the deformation parameter.

An alternative holographic description of \TTbar-deformed CFTs consists of imposing mixed boundary conditions at the asymptotic boundary~\cite{Guica:2019nzm}. In the absence of matter fields, this proposal has been shown to be equivalent to cutoff AdS for negative values of $\mu$. In contrast to cutoff AdS, mixed boundary conditions apply to both signs of the deformation parameter and furthermore allow the addition of matter fields in the bulk.  Nevertheless, the cutoff AdS proposal provides a simple geometric picture for the holographic correspondence that has been shown to reproduce several features of $T\bar T$-deformed CFTs and may furthermore shed light on the problem of bulk locality in AdS/CFT.

For completeness, let us mention that there is also a single-trace version of the $T\bar T$ deformation \cite{Giveon:2017nie}, which is defined as a symmetric product orbifold of $T\bar T$-deformed CFTs.\footnote{In order to distinguish the two versions of $T\bar T$, the usual \TTbar deformation is sometimes referred to as the double-trace version.} The bulk geometries that are dual to states in single-trace \TTbar-deformed CFTs are no longer locally AdS$_3$, but are spacetimes that can be generated by a TsT transformation \cite{Araujo:2018rho,Borsato:2018spz,Apolo:2019zai} (see also~\cite{Apolo:2018qpq,Apolo:2019yfj,Apolo:2021wcn} for earlier and related developments). The single-trace $T\bar T$ deformation provides another approach to holography for spacetimes that are not asymptotically AdS. Relatedly, there are other holographic proposals for \TTbar, such as the dual of the \TTbar-deformed MSW CFT~\cite{Maldacena:1997de} proposed in \cite{Manschot:2022lib}, and the de~Sitter version of $T\bar T$ holography proposed in \cite{Gorbenko:2018oov}. In this paper, we will not discuss these versions of holography and focus only on the original, double-trace version of the \TTbar deformation.

While the \TTbar deformation \eqref{TTbardefintro} can be defined for either sign of the deformation parameter $\mu$, the physical properties of the resulting theory depend significantly on this choice. In particular, when $\mu<0$, the energy of \TTbar-deformed CFTs becomes complex above some critical value. On the other hand, when $\mu>0$, the deformed theory features a Hagedorn growth of states at high energies, and a complex ground state energy if $\mu> \frac{3R^2}{c}$, where $c$ is the central charge of the undeformed CFT and $R$ is the radius of the cylinder. In particular, the torus partition function of $T\bar T$-deformed CFTs has been shown to be modular invariant in the range $0<\mu< \frac{3R^2}{c}$ \cite{Datta:2018thy}.

The sign of the deformation parameter also affects the conjectured holographic dualities. In particular, the cutoff AdS proposal for the double-trace \TTbar deformation is only valid when $\mu<0$, and there is so far no analogous proposal for positive values of $\mu$. In the single-trace version, the bulk geometry features closed timelike curves and curvature singularities when $\mu < 0$, in addition to a region in the spacetime where the Ricci scalar is positive \cite{Apolo:2019zai}. In contrast, the backgrounds with $\mu>0$ are free of the aforementioned pathologies except that the ground state solution features a complex dilaton when $\mu>\frac{3R^2}{c}$.

In this paper, we propose a \emph{glue-on} version of holography for double-trace \TTbar-deformed CFTs with $\mu>0$.  In this approach, instead of introducing a finite cutoff in the bulk, we glue a patch of an auxiliary AdS$_3$ spacetime to the asymptotic boundary of the original AdS$_3$ background. We can interpret this procedure as extending the cutoff surface from the original AdS$_3$ spacetime \emph{beyond} the asymptotic boundary where $\mu > 0$.\footnote{As we were completing the manuscript, the paper \cite{Kawamoto:2023wzj} appeared where two AdS spacetimes are glued at a finite radius and a $T\bar T$-like deformation arises from junction conditions. This is different from our prescription, where the gluing happens at the asymptotic boundary and the $T\bar T$-deformation emerges at a cutoff surface, which corresponds to the boundary of the glued geometry. AdS spacetimes glued at their asymptotic boundaries have been considered previously in the literature in other contexts, see e.g.~\cite{Kiritsis:2006hy,Aharony:2006hz,Apolo:2012gg}.} The glue-on AdS proposal can also be understood as a geometrization of the analytic continuation that takes $\mu \mapsto - \mu$ in cutoff AdS holography. This is manifest in the fact that the aforementioned auxiliary AdS$_3$ spacetime is obtained from the original AdS$_3$ background by analytic continuation of the radial coordinate. We will show that the glue-on AdS proposal is able to reproduce several features of $T\bar T$-deformed CFTs including the trace flow equation, the deformed energy spectrum, as well as the torus and sphere partition functions.

As described above, the double-trace $T\bar T$ deformation induces mixed boundary conditions at the asymptotic boundary which have been shown to be equivalent to cutoff AdS when $\mu < 0$. We will show that the mixed boundary conditions are also equivalent to glue-on AdS when $\mu > 0$, thereby providing a geometric picture for the correspondence for either sign of the deformation parameter.

The layout of this paper is as follows. In section \ref{se:cutoffholography} we review the \TTbar deformation and cutoff AdS holography for \TTbar-deformed CFTs with $\mu<0$. In section \ref{se:glueonproposal} we propose the glue-on picture of holography for \TTbar-deformed CFTs with $\mu>0$. In this section we show that the glue-on proposal reproduces the $T\bar T$ trace flow equation, its subluminal propagation, and the critical value of the deformation parameter. We provide additional evidence for the correspondence in section \ref{se:charges}, where we use the covariant formalism to compute the quasilocal energy in both the cutoff and glue-on cases, and find agreement with the spectrum of  \TTbar-deformed CFTs on the cylinder. Finally, in section \ref{se:partitionfunction} we compute the bulk on-shell action for Euclidean spaces with the topology of the torus and the sphere at the boundary, and show that they reproduce the partition functions of \TTbar-deformed CFTs on these geometries.


\section{The \TTbar deformation and cutoff AdS holography} \label{se:cutoffholography}

In this section we review the holographic correspondence between $T\bar T$-deformed CFTs with a negative deformation parameter and Einstein gravity on asymptotically AdS$_3$ spacetimes with a finite cutoff \cite{McGough:2016lol}.

\subsection{The \TTbar deformation} \label{se:ttbar}

The \TTbar deformation of a two-dimensional CFT can be defined via the following differential equation for the Lorentzian action~\cite{Zamolodchikov:2004ce,Smirnov:2016lqw,Cavaglia:2016oda},
\eq{ \label{TTbardef}
\partial_\mu I &= {8 \pi} \int d^2x \sqrt{-\gamma}\, T\bar T,
}
where $\mu$ is the deformation parameter, $\gamma_{ij}$ is the two-dimensional metric, and the \TTbar operator is defined by
\eq{
T\bar T = \frac{1}{8}\,\big( T^{ij}T_{ij}- (T^i_i)^2 \big), \qquad {T_{ij} = \frac{2}{\sqrt{-\gamma}} \frac{\d I}{\d \gamma^{ij}} }.
}
The stress tensor satisfies the conservation law $ \nabla_i  \langle T^{ij} \rangle = 0$ and the so-called trace flow equation \cite{McGough:2016lol,Shyam:2017znq}
\eq{
\langle T^i_i \rangle &= -\frac{c}{24\pi} R^{(2)}
+ 2\pi \mu \, \big (\langle T^{ij}T_{ij}\rangle-\langle T^i_i\rangle^2 \big). \label{floweq}
}
The expectation value of the $T\bar T$ operator factorizes and is a constant in any translationally invariant QFT, a fact that leads to a universal equation for the spectrum. In particular, the spectrum of $T\bar T$-deformed CFTs on a cylinder of size $2\pi R$ can be solved explicitly and the deformed energy and angular momentum can be shown to satisfy~\cite{Smirnov:2016lqw,Cavaglia:2016oda}
\eq{
E(\mu)&= - \frac{R }{ 2\mu } \bigg(1-\sqrt{1 + \frac{4\mu}{R} E(0) + \frac{4\mu^2}{R^4} J(0)^2 }
\,\bigg), \qquad J(\mu)=J(0),  \label{ttbarspectrum}
}
where $E(0)$ and $J(0)$ denote the undeformed values of the energy and angular momentum. In particular, the ground state energy can be obtained from \eqref{ttbarspectrum} by letting $E(0) = -c/12R$ and $J(0) = 0$, such that
\eq{
E_{\text{vac}}(\mu) = - \frac{R}{2 \mu} \bigg( 1 - \sqrt{ 1- \frac{ c\mu}{3 R^2}} \bigg).\label{Evac}
}

Crucially, the deformed theory depends on the sign of the deformation parameter $\mu$, as can be seen from the spectrum \eqref{ttbarspectrum}. When $\mu<0$, the spectrum becomes complex above a critical value of the energy
\eq{
E_c(\mu) = -\frac{R}{2 \mu}.
}
On the other hand, when $\mu>0$, the ground state energy \eqref{Evac} becomes complex unless the deformation parameter is smaller than the critical value
\eq{
\mu \le \mu_c \equiv \frac{3 R^2}{c}, \label{criticalmu}
}
where $c$ is the central charge of the undeformed CFT. In this case, the spectrum exhibits Hagedorn growth at high energies $E(\mu) \gg R/\mu$ and the log of the density of states scales as $\sqrt{c\mu/3}\,E(\mu)$. Relatedly, there is a restriction on the temperatures $T_{L,R}(\mu)$ conjugate to the deformed energies $E_{L,R}(\mu) \equiv \frac{1}{2}\big(E(\mu)\pm J(\mu)\big)$, which are bounded by \cite{Giveon:2017nie,Apolo:2019zai}
\eq{
T_L (\mu)\, T_R(\mu) \le T_H(\mu)^2 \equiv \frac{3}{4\pi^2 c\mu}, \label{productbound}
}
where $T_H(\mu)$ is the Hagedorn temperature.


\subsection{Cutoff AdS holography} \label{se:cutoffreview}

Let us consider a two-dimensional CFT that is holographically dual to three-dimensional Einstein gravity with a negative cosmological constant. For negative $\mu$, the \TTbar-deformation of such a CFT has been proposed to be dual to the same gravitational theory satisfying Dirichlet boundary conditions at a finite cutoff~\cite{McGough:2016lol}.

In order to describe the holographic dictionary, let us first discuss some aspects of the correspondence before the deformation. When $\mu = 0$, the bulk theory is described by Einstein gravity in asymptotically AdS$_3$ spacetimes satisfying Brown-Henneaux boundary conditions at the asymptotic boundary
\eq{
ds^2|_{\epsilon} = \frac{1}{\epsilon} ds_b^2, \qquad  ds_b^2 = \ell^2 du\, dv = \ell^2  (-dt^2 + d\vp^2),  \qquad \vp \sim \vp + 2\pi,
}
where $\epsilon \to 0$ is a UV cutoff, $\ell$ is the scale of AdS, and $ds_b^2$ is the line element at the boundary. The bulk lightcone coordinates $(u,v) = (\varphi + t, \varphi - t)$ are identified with the lightcone coordinates $(\hat x^+, \hat x^-)$ of the dual CFT such that
\eq{
(u,v) = \frac{1}{R}(\hat x^+, \hat x^-), \qquad R = \ell.
}

In three dimensions, the most general solution of pure Einstein gravity with Brown-Henneaux boundary conditions can be written as \cite{Banados:1998gg}
\eq{
ds^2 = \ell^2 \bigg( \frac{d\rho^2}{4 \rho^2} + \frac{ \big( du + \rho \, \mathcal {\bar L}(v)\, dv \big) \big( dv + \rho \, \mathcal L(u)\, du \big) }{\rho} \bigg) , \qquad \rho > 0, \label{banados}
}
where $\mathcal L(u)$ and $\bar{\mathcal L}(v)$ are arbitrary periodic functions of their arguments. In particular, the global AdS$_3$ vacuum and the BTZ black hole are characterized by constant values of $\mathcal L(u)$ and $\bar{\mathcal L}(v)$ that can be parametrized by
\eq{
\mathcal L(u) =T_u^2, \qquad \mathcal {\bar L}(v) = T_v^2. \label{btzparam}
}
For  $T_u = T_v = i/2$, the metric \eqref{banados} describes the AdS$_3$ vacuum, while for $T_u \ge 0$ and $T_v \ge 0$, it describes BTZ black holes with left and right-moving temperatures respectively given by $T_u/\ell\pi$ and $T_v/\ell\pi$. In this gauge the asymptotic boundary is located at $\rho = \epsilon \to 0$.

According to the cutoff AdS proposal \cite{McGough:2016lol}, the boundary of the asymptotically AdS$_3$ spacetimes \eqref{banados} is moved into the interior of the spacetime after the deformation. Since the undeformed CFT is defined on a cylinder of size $2\pi R$, the cutoff surface where the deformed theory is defined must also be a cylinder with the same periodic identification. Furthermore, we note that the proper length of the $\vp$-circle naturally introduces a scale which can be identified with the cutoff scale.  Let us define the dimensionless radial coordinate $\zeta$ by the size of the $\vp$-circle. Then, the cutoff surface can be defined via
\eq{
\frac{1}{\zeta_c} = \frac{1}{\zeta(x)} \equiv \ell^{-2} g_{\varphi\varphi},  \label{cutoff}
}
where $g_{\mu\nu}$ denotes the bulk metric and $\zeta_c$ is a constant cutoff scale. In particular, for the solutions \eqref{banados} the cutoff surface is located at
\eq{
\rho_c(u,v) = \frac{2}{ \zeta_c^{-1} - \mathcal L(u) - \mathcal{\bar L}(v) + \sqrt{\big[\zeta_c^{-1} - \mathcal L(u) - \mathcal{\bar L}(v)   \big]^2 - 4 \mathcal L(u)\,\mathcal{\bar L}(v)}}, \ \quad \rho_c > 0. \label{rhocn}
}
Note that in this gauge the radial coordinate $\rho$ is generically not constant at the cutoff surface, which depends on the solution-specifying functions $\mathcal L(u) $ and $ \mathcal{\bar L}(v)$, in analogy with the boundary conditions imposed in JT gravity \cite{Maldacena:2016upp}.

It is convenient to write the induced metric at the cutoff surface in locally Cartesian coordinates without changing the identification of the spatial circle, namely
\eq{
ds^2\big|_{\zeta_c} \equiv \frac{1}{\zeta_c} ds_c^2, \qquad ds_c^2 = \ell^2 du' dv' = \ell^2 ( -dt'^2 + d\vp'^2) , \qquad \vp' \sim \vp' + 2\pi,
\label{locallyflat}
}
where $ds_c^2$ denotes the line element at the cutoff surface. In this gauge, Dirichlet boundary conditions at the cutoff surface are implemented by requiring~\cite{McGough:2016lol,Kraus:2018xrn,Hartman:2018tkw}
\eq{
\delta g_{\mu\nu}'(x'^\mu) \big|_{\zeta_c} = 0. \label{eq:dirichlet}
}
A general solution can then be obtained by expanding the metric near the cutoff surface and solving Einstein's equations such that
\eq{
ds^2 = g_{\mu\nu}' dx'^{\mu} dx'^{\nu} = g_{\mu\nu}'\big|_{\zeta_c}\,\, dx'^{\mu} dx'^{\nu} + \mathcal O (\d \zeta),
}
where $\d \zeta$ denotes the radial distance from the cutoff. In particular, any solution satisfying Brown-Henneaux boundary conditions at the asymptotic boundary before the deformation can be put into the locally Cartesian coordinates \eqref{locallyflat} after the deformation, whereupon it satisfies Dirichlet boundary conditions at the cutoff surface.

The holographic dictionary identifies the primed coordinates $(u', v')$ with the lightcone coordinates $(x^+, x^-)$ of the dual \TTbar-deformed CFT, namely
\eq{
(u',v') = \frac{1}{\ell}(x^+, x^-).
}
Consequently, the dual field theory can be thought of as living at the cutoff surface with line element $ds^2_c$ given in \eqref{locallyflat}. Crucially, the cutoff radius $\zeta_c$ is related to the deformation parameter $\mu$ in a universal way by \cite{McGough:2016lol}
\eq{
\zeta_c = - \frac{c \mu}{3\ell^2}. \label{dictionary}
}
Note that when $\mu \to 0$, the cutoff radius is pushed to the asymptotic boundary such that $\zeta_c \to 0$, $(u',v') \to (u,v)$, and we recover the standard dictionary of the AdS/CFT correspondence.

The cutoff proposal is valid only for negative values of the deformation parameter, which is necessary for the cutoff radius to be positive. In the next section we will show that an analog of the cutoff picture exists for positive values of the deformation parameter. In this case, instead of a finite cutoff in the bulk, we propose a finite cutoff on an auxiliary AdS$_3$ spacetime that is glued to the original AdS$_3$ background at the asymptotic boundary.


\section{Glue-on AdS holography} \label{se:glueonproposal}

In this section we propose that cutoff AdS holography can be extended to positive values of the deformation parameter by gluing an auxiliary AdS$_3$ spacetime at the asymptotic boundary. In this case, the cutoff surface is pushed beyond the asymptotic boundary of the original spacetime and lies on the auxiliary glue-on geometry. We show that the proposal reproduces several features of $T\bar T$-deformed CFTs with a positive deformation parameter, including the trace flow equation, the critical value of the deformation parameter, and subluminal propagation. We also comment on the relationship between the glue-on AdS proposal and Einstein gravity with mixed boundary conditions at the asymptotic boundary.

\subsection{Motivating example} \label{se:example}

In order to motivate our proposal let us first consider the massless BTZ black hole
\eq{\label{masslessBTZ}
\textrm{BTZ}: \quad ds^2 &= \ell^2 \bigg( \frac{d\rho^2}{4\rho^2} + \frac{du\,dv}{\rho} \bigg), \qquad \rho\ge 0,
}
where $(u,\,v) \sim (u+2\pi,\,v+2\pi)$. In this case, the cutoff surface is found via \eqref{cutoff} to be located at
\eq{
\rho_c  = \zeta_c= - \frac{c \mu}{3 \ell^2}. \label{rhocmasslessbtz}
}
We see that when $\mu$ is negative, the cutoff surface lies in the interior of \eqref{masslessBTZ} so that the cutoff surface is pushed toward the asymptotic boundary as $\mu \to 0^-$.

On the other hand, when $\mu$ is positive the surface defined by \eqref{rhocmasslessbtz} leads to a negative value of $\rho_c$. This value of the radial coordinate is not supported in locally AdS$_3$ spacetimes and hence the original cutoff AdS proposal does not apply. In order to describe the holographic correspondence for $\mu > 0$, we first introduce an auxiliary, locally AdS$_3$ spacetime dubbed BTZ$^*$ that is given by
\eq{
\textrm{BTZ}^*\!: \quad ds^2 &= \ell^2 \bigg( \frac{d\rho^2}{4\rho^2} + \frac{du\,dv}{\rho} \bigg),  \qquad  \rho  < 0, \label{masslessBTZstar}
}
and satisfies the same identification of coordinates as \eqref{masslessBTZ}. This spacetime can be thought of as a mirror image of the massless BTZ black hole where the $t$ and $\vp$ coordinates exchange roles in the sense that the noncompact direction along $\p_t$ becomes spacelike while the circle along $\p_\vp$ becomes timelike. In analogy with BTZ, the $\rho \to -\infty$ surface does not correspond to the boundary of BTZ$^*$ but to the image of the (massless) BTZ horizon.

Let us now glue the asymptotic boundaries of BTZ and BTZ$^*$ together so that the union BTZ $\cup$ BTZ$^*$ is described by the same metric \eqref{masslessBTZ} but with the range of the radial coordinate extended to the entire real axis, namely\footnote{This can be made more precise by introducing a cutoff $\rho=\epsilon$ at the asymptotic boundary  of BTZ, a cutoff $\rho=-\epsilon$ at the asymptotic boundary of BTZ$^*$, and then identifying the two hypersurfaces, as described in more detail shortly.}
\eq{
\text{BTZ} \cup \textrm{BTZ}^*\!: \quad ds^2  = \ell^2 \bigg( \frac{d\rho^2}{4\rho^2} + \frac{du\,dv}{\rho} \bigg), \qquad   \rho  \in \mathbb R. \label{masslessBTZ2}
}
The cutoff AdS proposal can be extended to all values of the $T\bar T$ deformation parameter using \eqref{rhocmasslessbtz} such that the cutoff surface lies in either the original ($\rho > 0$) region of \eqref{masslessBTZ2} when $\mu < 0$, or in the glue-on ($\rho < 0$) region when $\mu > 0$. In the next section we  provide a more rigorous formulation of the proposal and show how the construction above can be generalized to any locally AdS$_3$ spacetime.


\subsection{Glue-on AdS and \TTbar with \texorpdfstring{\ensuremath{\mu > 0}}{mu > 0}} \label{se:glueondef}
\begin{figure}[!ht]
\centering
\includegraphics[scale=0.5]{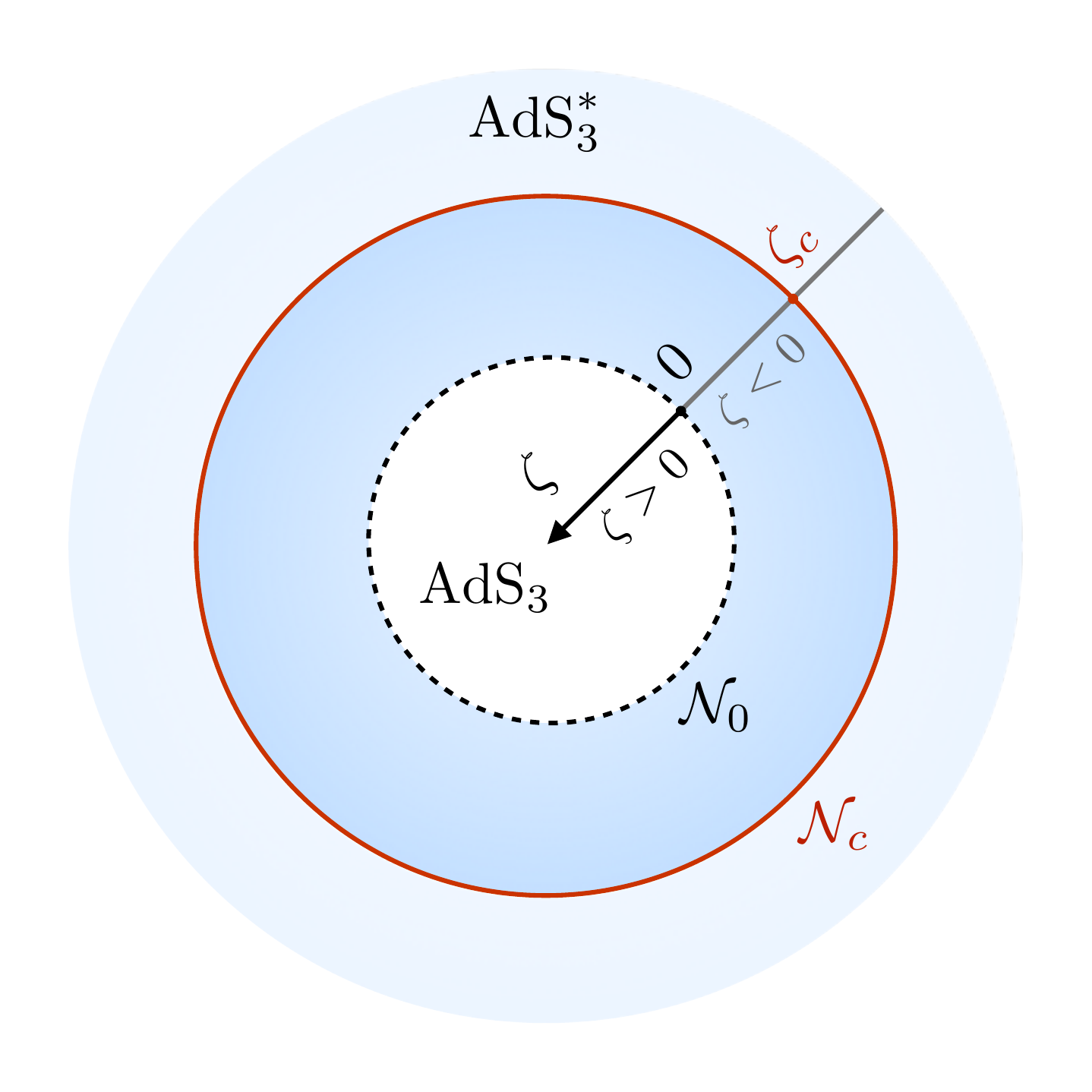}
\caption{Illustration of an extended AdS$_3$ $\cup$ AdS$_3^*$ spacetime: a fixed-$t$ slice  of an AdS$_3$ spacetime (white disk) is glued to the fixed-$t$ slice of an auxiliary AdS$_3^*$ spacetime (blue annulus) along their asymptotic boundaries $\mathcal N_0$ at $\zeta \to 0$ (dashed circle). When $\mu > 0$, the cutoff surface $\mathcal N_c$ at $\zeta = \zeta_c$ (red circle) lies on the auxiliary AdS$_3^*$ part of the extended AdS$_3$ spacetime.}\label{glueon}
\end{figure}

Let us consider an extended AdS$_3$ $\cup$ AdS$_3^*$ spacetime obtained by gluing the asymptotic boundaries of two locally AdS$_3$ spacetimes related by analytic continuation. We propose that the holographic description of $T\bar T$-deformed CFTs with $\mu > 0$ is equivalent to introducing a cutoff in the auxiliary AdS$_3^*$ part of spacetime. The resulting bulk geometry will be referred to as glue-on AdS$_3$, as it involves gluing a patch of a locally AdS$_3$ spacetime to the original AdS$_3$ background. See fig.~\ref{glueon} for an illustration.

In order to construct the glue-on geometry associated with a locally AdS$_3$ spacetime, let us first foliate this spacetime by a one-parameter family of timelike surfaces ${\mathcal N}_\zeta$ that are defined in terms of a radial function $\zeta (x^\mu)$ by
\eq{
{\mathcal N}_\zeta\,\colon\quad \zeta(x^\mu) = \textsl{constant}. \label{cutoffg}
}
In terms of the outward-pointing unit normal vector $n^\mu \propto \nabla^\mu \zeta$, the AdS$_3$ metric can be decomposed as
\eq{
\text{AdS}_3: \quad ds^2= \frac{1}{ \zeta} \gamma_{ij}dx^i dx^j+n_\mu n_\nu dx^\mu dx^\nu, \qquad \zeta> 0,\label{AdS3metric}
}
where $x^i, \, i = \{0,1\}$ are the coordinates on the hypersurface $\mathcal N_{\zeta}$ and $\frac{1}{\zeta}  \gamma_{ij}$ is the induced metric on $\mathcal N_{\zeta}$. The original locally AdS$_3$ spacetime is parameterized by the coordinates $(x^i, \zeta)$ where $\zeta > 0$ and the asymptotic boundary is located at $\zeta \to 0^+$. In analogy with the massless BTZ$^*$ background considered in the previous section, we can construct an auxiliary AdS$_3^*$ spacetime by analytically continuing \eqref{AdS3metric} to negative values of $\zeta$, such that the extended AdS$_3$ $\cup$ AdS$_3^*$ geometry is described by the metric \eqref{AdS3metric} but with $\zeta\in \mathbb R$.  In this construction, we have to be careful about the boundary conditions imposed at the asymptotic boundary as
$\zeta\to 0^+$ and  $\zeta\to 0^-$. More precisely, we first have to introduce near-boundary surfaces $\mathcal N_{\zeta={ \epsilon}}$ in the original AdS$_3$ region and $\mathcal N_{\zeta=-{ \epsilon}}$ in the auxilliary AdS$_3^*$ region,  glue them together, and finally send $\epsilon \to 0.$ We formally denote the boundary surface by the limit $\mathcal N_{0}=\mathcal N_{0^+}=\mathcal N_{0^-}$, though we need to keep track of the asymptotic cutoff $\epsilon$ in actual computations.

In the AdS$_3$ $\cup$ AdS$_3^*$ spacetime, the cutoff surface for either sign of $\mu$ can be uniformly defined as
\eq{
{\mathcal N}_c\,\colon\quad \zeta(x^\mu) = \zeta_c\equiv -\frac{c\mu}{3\ell^2}, \label{generalcutoff}
}
where we must impose Dirichlet boundary conditions for the metric. We then propose that the cutoff $(\zeta_c > 0)$ and glue-on $(\zeta_c < 0)$ spacetimes
\eq{
&\text{cutoff}/\text{glue-on AdS}_3: \quad ds^2= \frac{1}{\zeta} \gamma_{ij}dx^i dx^j+n_\mu n_\nu dx^\mu dx^\nu, \qquad \zeta\ge \zeta_c \in \mathbb R, \label{glueonAdS}
}
are holographically dual to states in $T\bar T$-deformed CFTs living on the manifold ${\mathcal N}_c$ with metric $\gamma_{ij}$.
In particular, cutoff AdS$_3$ is the region of AdS$_3$ satisfying $\zeta \ge \zeta_c>0$. On the other hand, the proposed glue-on AdS$_3$ spacetime, which consists of $\zeta \ge \zeta_c$ with $\zeta_c< 0$, is obtained by gluing the patch of auxiliary AdS$_3^*$ lying between the surfaces $\mathcal{N}_c$ and ${\mathcal N}_{0^-}$ to the original AdS$_3$ spacetime (see fig.~\ref{glueon}). Note that in both the cutoff and glue-on cases, it is the rescaled metric $\gamma_{ij}$ in \eqref{glueonAdS}, instead of the induced metric, that is identified with the metric of the \TTbar-deformed CFT.

Let us now show how the glue-on AdS proposal \eqref{glueonAdS} and Einstein's equations can be used to reproduce one of the characteristic features of $T\bar T$-deformed CFTs, namely the trace flow equation \eqref{floweq}. The argument is a straightforward generalization of the one used in the cutoff case \cite{Kraus:2018xrn,Kraus:2021cwf}. Indeed, using the Gauss-Codazzi equations and  \eqref{glueonAdS}, the normal-normal component and the normal-tangential components of Einstein's equations without matter at $\zeta = \zeta_c$ can be respectively written as
\eq{
0&= \zeta_c R^{(2)} + \zeta^2_c \Big(K^{ij}K_{ij} - K^2\Big) + \frac{2}{\ell^2}, \label{nn}\\
0&=\nabla_i(K^{ij}-\gamma^{ij} K), \label{nt}
}
where $K_{ij} = \frac{1}{2} \mathcal{L}_{n} g_{ij}$ is the extrinsic curvature, $R^{(2)}$ is the Ricci scalar of $\gamma_{ij}$, and indices are raised/lowered using the rescaled induced metric $\gamma_{ij}$ on the cutoff surface \eqref{cutoffg}. The Brown-York stress tensor conjugate to $\gamma_{ij}$ at the cutoff surface is given by \cite{Balasubramanian:1999re}
\eq{
T_{ij}= \frac{\sigma_{\zeta_c}}{8\pi G} \Big( K_{ij} -  K \gamma_{ij} + \frac{1}{\ell |\zeta_c|}\gamma_{ij}\Big), \qquad \sigma_{\zeta_c} = \frac{\abs{\zeta_c}}{\zeta_c} = \pm 1,
\label{BYtensor}
}
and is identified with the stress tensor of the dual $T\bar T$-deformed CFT.
The $\sigma_{\zeta_c}$ factor is a result of conjugation with respect to the boundary metric $\gamma_{ij}$, along with our choice of the normal vector $n^\mu$, which points from the bulk toward the cutoff surface $\mathcal{N}_{\zeta_c}$, i.e.~in the $-\zeta$ direction.\footnote{More precisely, the Brown-York stress tensor conjugate to the induced metric $h_{ij} \equiv \frac{1}{\zeta} \gamma_{ij}$ is given by ${T^h_{ij} = \sigma_{\zeta_c} T_{ij}}$, such that $\sqrt{-\gamma}\,T_{ij} \var{\gamma^{ij}} = \sqrt{-h}\,T^h_{ij} \var{h^{ij}}$.}

In terms of \eqref{BYtensor}, the normal-tangential components of Einstein's equations \eqref{nt} reduce to the conservation law of the stress tensor $\nabla_i  T^{ij} = 0$. Furthermore, the normal-normal component \eqref{nn} reproduces the trace flow equation of $T\bar T$-deformed CFTs \eqref{floweq} provided that the cutoff scale $\zeta_c$ and the deformation parameter $\mu$ are related by the holographic dictionary \eqref{generalcutoff}. Note that the matching between Einstein's equations and the trace flow equation is valid for both signs of $\mu$, and hence it is valid for both the cutoff and glue-on proposals. Furthermore, note that the discussion in this section is general, and applies in principle to any choice of radial function \eqref{cutoffg}, as well as to any topology and metric $\gamma_{ij}$ on $\mathcal{N}_\zeta$.


\subsection{Comments on mixed boundary conditions}

Before we provide additional evidence for the glue-on AdS proposal, let us first consider its relationship to the mixed boundary conditions of \cite{Guica:2019nzm}.

In the AdS/CFT correspondence, deformations of the CFT by double-trace operators lead to a change in the boundary conditions of the bulk dual fields \cite{Klebanov:1999tb,Witten:2001ua}. A similar phenomenon is observed for the $T\bar T$-deformation, which induces mixed boundary conditions at the asymptotic boundary for either sign of the deformation parameter \cite{Guica:2019nzm}. Let us parametrize a general solution to the equations of motion of pure Einstein gravity by
\eq{
ds^2 = \ell^2 \frac{d\rho^2}{4 \rho^2} + \pqty{\frac{1}{\rho} g^{(0)}_{ij} + g^{(2)}_{ij} + \rho\thinmspace g^{(4)}_{ij}}  dx^i dx^j, \label{fggauge}
}
where $g^{(0)}_{ij}$ is the boundary metric. The subleading component $g^{(2)}_{ij}$ is related to the Brown-York stress tensor $T_{ij}$ and the Ricci scalar $\mathcal{R}[g^{(0)}]$ of the boundary metric via
\eq{
T_{ij} = \frac{1}{8\pi G\ell} \biggl( -\frac{\ell^2}{2} \, g^{(0)}_{ij} \mathcal{R}\big[g^{(0)}\big] - g^{(2)}_{ij} \biggr),
}
while $g^{(4)}_{ij}$ is determined from Einstein equations by
\eq{
g^{(4)}_{ij} = \frac{1}{4} g^{(2)}_{ik} g^{(0)\thinmspace kl}_{\phantom{kl}} g^{(2)}_{lj}.
}
When the boundary metric is flat, the solution \eqref{fggauge} reduces to \eqref{banados} where the nonvanishing components of the stress tensor are $T_{uu} = -\frac{\ell}{8\pi G} \mathcal L(u)$ and $T_{vv} = -\frac{\ell}{8\pi G} \bar{\mathcal L}(v)$.

The mixed boundary conditions induced by the $T\bar T$ deformation fix the following combination of leading and subleading components of the metric \cite{Guica:2019nzm}
\eq{
\delta \bigg( g^{(0)}_{ij} - \frac{\mu}{2 G \ell}\, g_{ij}^{(2)} + \Big(\frac{\mu}{2 G \ell} \Big)^2 g_{ij}^{(4)} \bigg) = 0. \label{mixedbc}
}
When  $\mu \to 0$, the mixed boundary conditions reduce to the standard Dirichlet boundary conditions that fix only the leading $g^{(0)}_{ij}$ component of the bulk metric. As noted in \cite{Guica:2019nzm}, when $\mu < 0$ the mixed boundary conditions \eqref{mixedbc} are equivalent to Dirichlet boundary conditions on the induced metric at the cutoff surface $\rho = \zeta_c$ such that
\eq{
\delta g_{ij} \big|_{\zeta_c} = \frac{1}{\zeta_c} \delta \bigg( g^{(0)}_{ij} + \zeta_c\, g_{ij}^{(2)} + \zeta_c^2\, g_{ij}^{(4)} \bigg) = 0, \qquad \zeta_c  = - \frac{\mu c}{3\ell^2}. \label{cutoffmixed}
}
Thus, in the absence of matter fields, the mixed boundary conditions \eqref{mixedbc} induced by the $T\bar T$ deformation with $\mu < 0$ are equivalent to the Dirichlet boundary conditions of the cutoff AdS proposal.\footnote{In the presence of matter fields the normal-normal component of Einstein's equations \eqref{nn} receives additional contributions that spoil the trace flow equation at the cutoff surface.
} Note that \eqref{cutoffmixed} continues to be meaningful for positive values of $\mu$ in an extended version of \eqref{fggauge} where $\rho \in \mathbb R$. As a result, the mixed boundary conditions \eqref{mixedbc} with $\mu > 0$ are also equivalent to the Dirichlet boundary conditions of the glue-on AdS version of holography proposed in this paper.

It is important to note that the relationship between the radial cutoff and the deformation parameter in \eqref{cutoffmixed} differs from the one used in the cutoff picture \eqref{rhocn} and also in the glue-on picture discussed momentarily in \eqref{cutoff-glueon}. This can be understood as different choices of the radial function $\zeta$ in \eqref{cutoffg}. The derivation of \eqref{mixedbc} corresponds to the choice $\zeta = \rho$, and consequently the cutoff in the $\rho$ coordinate is independent of the phase space variables. On the other hand, from \eqref{cutoffmixed} we learn that the metric and the size of the spatial circle at the cutoff surface generically depend on the phase space variables, which is different from the spatial circle used at the asymptotic boundary, and complicates the comparison between the deformed and undeformed theories. In contrast, the choice \eqref{cutoff} guarantees that the spatial circle is the same both at the asymptotic boundary and at the cutoff surface, which makes the aforementioned comparison straightforward. 

 The two radial functions discussed above are the same for the massless BTZ black hole since the subleading components of the metric vanish in this case. For other backgrounds, the two radial functions differ from each other, but the results at fixed $\mu$ reproduce the same results provided that we account for the aforementioned differences in the size of the spatial circle \cite{Guica:2019nzm}. More generally, we can consider the space of $T\bar T$-deformed CFTs with all allowed values of the deformation parameter $\mu$ and cylinder radius $R$. Different choices of the radial function correspond to different trajectories in the $(\mu,\, R)$ plane as $\mu$ is varied. Nevertheless, at a given point in the $(\mu,\, R)$ plane, physical observables obtained from either the mixed boundary conditions or the glue-on AdS proposal can be matched with each other. Note that our proposal and the derivation of the flow equation using Einstein's equations \eqref{nn} admits arbitrary choices of the radial function, and is therefore compatible with both cutoff AdS and mixed boundary conditions.


\subsection{Implications of the proposal} \label{se:evidence}

Let us now consider the implications of the glue-on proposal to $T\bar T$-deformed CFTs on a cylinder of size $2\pi \ell$. In the bulk, the general solution to pure Einstein gravity with Brown-Henneaux boundary conditions is given in \eqref{banados}, where the lightcone coordinates satisfy the identification $(u, v) \sim (u + 2\pi, v + 2\pi)$. In order to obtain the extended AdS$_3$ $\cup$ AdS$_3^*$ spacetime, we first need to choose a radial function \eqref{cutoffg}. As discussed in the previous section, there are two natural choices. In this section we stick to the choice of the cutoff AdS proposal  \eqref{cutoff}, the latter of which keeps the size of the cylinder fixed so that the field theory couples to the same background metric along the $T\bar T$ flow. The advantage of this choice is that the spectrum of the deformed and undeformed theories can be directly compared. In our conventions, the radial function can be written as
\eq{
\frac{1}{\zeta}\equiv\ell^{-2}g_{\varphi\varphi}, \label{choicezeta}
}
then the glue-on AdS$_3$ spacetime is given by \eqref{banados}, but with an extended range of coordinates satisfying $\zeta\ge \zeta_c\equiv -\frac{c\mu}{3\ell^2}$. We focus on the patch of \eqref{banados} that is connected to the asymptotic boundary, and has no other singularities, which means that
 \eq{
 	1 - \rho^2 \mathcal L(u)\,\mathcal{\bar L}(v) \ge 0.
 	\label{eq:rhoBound}
 }
 In the $(u,\,v,\,\rho)$ coordinates, the fixed-$\zeta$ surfaces are coordinate dependent as a result of inverting  \eqref{choicezeta}, and satisfy
\eq{
\rho= \frac{2\zeta}{1 - \zeta \big(\mathcal L(u) + \mathcal{\bar L}(v)\big) + \sqrt{\big[ 1 - \zeta \big(\mathcal L(u) + \mathcal{\bar L}(v)\big) \big]^2 - 4 \zeta^2 \mathcal L(u)\,\mathcal{\bar L}(v)}}, \label{rhoc}
}
where $\mathcal L(u)$ and $\mathcal{\bar L}(v)$ are periodic functions of $\varphi$. We see that the coordinate $\rho$ can take negative values when $\zeta<0$. In addition, we note that due to the periodic identification $(u, v) \sim (u + 2\pi, v + 2\pi)$, the AdS$_3^*$ region of the spacetime features a compact timelike coordinate $\vp$ while the noncompact coordinate $t$ becomes spacelike. Nevertheless, whether the cutoff surface is located on the AdS$_3$ or AdS$_3^*$ regions of the spacetime, we interpret the dual $T\bar T$-deformed CFT as living on a surface with a spacelike circle whose line element $ds_c^2$ is determined from the induced metric at the cutoff surface via \eqref{locallyflat}, namely
\eq{
ds^2_c = \ell^2 du' dv' = \ell^2 ( -dt'^2 + d\vp'^2) , \qquad \vp' \sim \vp' + 2\pi.\label{cutoffmetric}
}

Let us now focus on the zero mode backgrounds obtained from \eqref{banados} by letting $\mathcal L(u) =T_u^2$ and $\mathcal {\bar L}(v) = T_v^2$. In this case, a negative value of $\zeta$ corresponds to a negative value of $\rho$, and vice versa. The corresponding AdS$_3$ $\cup$ AdS$_3^*$ spacetimes read
\eq{
ds^2 = \ell^2 \bigg( \frac{d\rho^2}{4 \rho^2} + \frac{ \big( du + \rho \, T_v^2 dv \big) \big( dv + \rho \, T_u^2 du \big) }{\rho} \bigg), \qquad  \rho \in \mathbb R. \label{extendedbtz}
}
Note that for the extended BTZ black holes with $T_u \ge 0$ and $T_v \ge 0$, the auxiliary $\rho < 0$ region of the spacetime is not geodesically complete. In this case, the AdS$_3^*$ part of the spacetime features a horizon at $\rho = \rho_h^* \equiv -1/T_u T_v$ where the Killing vector $T_u^{-1}\p_u + T_v^{-1} \p_v$ becomes degenerate. In particular, for the non-rotating case where $T_u = T_v = \pi/\beta_t$, the Killing vector is $\frac{\beta_t}{\pi} \pdd{\varphi}$. The surface $\rho = \rho_h^*$ can be thought of as the mirror image of the horizon of the BTZ black hole which is located at $\rho = \rho_h \equiv 1/T_u T_v$. Furthermore, note that \eqref{eq:rhoBound} corresponds to the region between the horizons, namely $\rho_h^* \le \rho \le \rho_h$.

Let us now show that our proposal of extending the cutoff surface beyond the asymptotic boundary reproduces several features of $T\bar T$-deformed CFTs with $\mu > 0$. According to the cutoff/glue-on AdS proposal, we interpret the effect of the \TTbar deformation as moving the asymptotic boundary of the extended AdS$_3$ spacetimes \eqref{extendedbtz} from $\rho = 0$ to the interior of either the original ($\rho > 0$) or the auxiliary ($\rho < 0$) parts of the spacetime. From \eqref{rhoc}, we see that the cutoff surface is located at a constant value of the radial coordinate $\rho_c$ that is given by
\eq{
\rho_c = \frac{2\zeta_c}{ 1 - \zeta_c (T_u^2 + T_v^2) + \sqrt{ \big(1 - \zeta_c (T_u + T_v)^2 \big) \big(1 - \zeta_c (T_u - T_v)^2 \big) } }.\label{cutoff-glueon}
}
The location of the cutoff surface is real provided that
\eq{
\big(1 - \zeta_c (T_u+T_v)^2 \big) \big(1 - \zeta_c (T_u-T_v)^2 \big) \ge 0. \label{cond1}
}

\paragraph{Extended BTZ.}

For real values of $T_u$ and $T_v$, the condition \eqref{cond1} is always satisfied when $\mu > 0$ ($\zeta_c < 0$), in which case $\rho_c < 0$ and the cutoff surface lies in the auxiliary part of the spacetime. In this case, we find that $\rho_c$ satisfies the following lower bound
\eq{
\rho_c > - \textrm{min} \Big(\frac{1}{T_u^2}, \frac{1}{T_v^2}\Big)\ge - \frac{1}{T_u T_v} = \rho^*_h, \label{lowerbound}
}
where the minimum value of $\rho_c$ is approached as $\mu \to \infty$. The second inequality in \eqref{lowerbound} tells us that the surface \eqref{cutoff-glueon} lies between the asymptotic boundary and the horizon $\rho_h^*$ on the auxiliary side of the geometry.

On the other hand, when $\mu < 0$ ($r_c^2 > 0$), the cutoff surface lies in the $\rho > 0$ region of the spacetime and \eqref{cond1} is the same condition that guarantees a real $T\bar T$ spectrum, as described in more detail in  section \ref{se:charges}. In this case, \eqref{cutoff-glueon} is real and positive provided that
\eq{
\zeta_c  \le \frac{1}{(T_u + T_v)^2}.
}
This condition tells us that as $\mu$ is decreased, the cutoff surface is moved towards the interior of the spacetime until it reaches the outer horizon of the BTZ black hole at $\rho = \rho_h$.

\paragraph{Extended global AdS\textsubscript{3}.}

As a special case of the zero mode backgrounds let us consider the extended version of the global AdS$_3$ vacuum where $T_u = T_v = i/2$. In this case the cutoff surface is located at
\eq{
\rho_c = \frac{2\zeta_c}{ 1 + \sqrt{1+ \zeta_c} + \zeta_c/2 }. \label{rhocvacuum}
}
In contrast to the general case with $T_{u, v} \in \mathbb R$, we see that $\rho_c$ is always real when $\zeta_c > 0$ so that $\mu$ can become arbitrarily negative. Furthermore, we see that \eqref{rhocvacuum} is also real when
\eq{
-1 \le \zeta_c < 0 \quad \Leftrightarrow \quad 0 < \mu\le \frac{ 3\ell^2 }{c}. \label{reality}
}
Note that the upper bound on $\mu$ reproduces the critical value of the deformation parameter in $T\bar T$-deformed CFTs \eqref{criticalmu}.

\paragraph{State-dependent map of coordinates.}

At the cutoff surface  \eqref{cutoff-glueon}, we must impose Dirichlet boundary conditions such that the induced metric is given by \eqref{locallyflat}. This can be accomplished by the state-dependent change of coordinates \cite{McGough:2016lol}\footnote{For the extended version of the locally AdS$_3$ spacetimes \eqref{banados}, the change of coordinates \eqref{cc1} and \eqref{cc2} is nonlocal with $dt' = \sqrt{[ 1 - \zeta_c \big( \mathcal L(u) + \mathcal{\bar L}(v) \big) ]^2 - 4 \displaystyle \zeta_c^2 \mathcal L(u)\,\mathcal{\bar L}(v)}\,dt$ and $d\vp' = d\vp + \zeta_c \big[ \mathcal L(u) - \bar{\mathcal L}(v) \big] dt$.}
\eq{
dt' &= \sqrt{ \big(1 - \zeta_c (T_u+T_v)^2 \big) \big(1 - \zeta_c (T_u-T_v)^2 \big) } \, dt,  \label{cc1} \\
 d\vp' &= d\vp + \zeta_c \big( T_u^2 - T_v^2 \big)\,dt. \label{cc2}
}
The change of coordinates \eqref{cc1} and \eqref{cc2} are valid for both signs of $\mu$, i.e.~they are valid for cutoff surfaces in either the AdS$_3$ or AdS$_3^*$ parts of the spacetime. For the zero mode backgrounds considered in this section, the change of coordinates \eqref{cc1} and \eqref{cc2} is local and the square-root structure is important for reproducing the spectrum of \TTbar-deformed CFTs as we will see in section \ref{se:charges}.

The change of coordinates \eqref{cc1} and \eqref{cc2} allow us to relate deformed quantities in $T\bar T$-deformed CFTs to undeformed ones. In particular, we note that for the zero mode backgrounds considered herein a null perturbation at the asymptotic boundary with $v_\pm \equiv \pm d\vp/dt = 1$ is seen from the cutoff surface to satisfy
\eq{
v'_{\pm} \equiv \pm \frac{d\vp'}{dt'} = \frac{ 1 \pm \zeta_c (T_u^2 - T_v^2)  }{\sqrt{ \big(1 - \zeta_c (T_u+T_v)^2 \big) \big(1 - \zeta_c (T_u-T_v)^2 \big) }}.
}
Expanding this expression around $\mu = 0$ we obtain
\eq{
v'_{+} &\sim 1 + 2 \zeta_c T_{u}^2 + \mathcal O(\zeta_c^2) = 1 - \frac{4\mu}{\ell} E_L + \mathcal O(\mu^2), \label{speeds}
}
where $E_L = (c/6 \ell)\, T_u^2 > 0$ denotes the left moving energy in the undeformed CFT. A similar expression for $v'_-$ can be obtained  by letting $u \leftrightarrow v$ in \eqref{speeds}. Note that when $\mu < 0$, null perturbations at the asymptotic boundary propagate superluminally at the cutoff surface, in agreement with \TTbar-deformed CFTs with a negative deformation parameter, as observed in \cite{McGough:2016lol,Marolf:2012dr}. On the other hand, when $\mu > 0$, null perturbations propagate subluminally at the cutoff surface with respect to the line element \eqref{cutoffmetric}, as expected for \TTbar-deformed CFTs with positive $\mu$.

\bigskip

To summarize, we have proposed that the holographic description of $T\bar T$-deformed CFTs with a positive deformation parameter is equivalent to pushing the boundary of AdS$_3$ beyond the asymptotic boundary to an auxiliary AdS$_3^*$ spacetime. Using the glue-on proposal, we have reproduced the critical value of the deformation parameter \eqref{criticalmu}, the subluminal propagation of light-like signals \eqref{speeds}, and the trace flow equation \eqref{floweq} of \TTbar deformed CFTs with $\mu>0$. Further evidence for our proposal will be presented in section \ref{se:charges}, where we compute the quasilocal energy of BTZ black holes for both signs of the deformation parameter, the latter of which will be shown to agree with the spectrum of $T\bar T$-deformed CFTs. In addition, in section \ref{se:partitionfunction} we use the cutoff/glue-on AdS proposals to compute the partition function of $T\bar T$-deformed CFTs on the torus and the sphere, expressions that will be shown to match the partition functions computed from field theory.


\section{Quasilocal charges in the covariant formalism} \label{se:charges}

In this section, we revisit the calculation of the quasilocal energy of BTZ black holes using the covariant formalism \cite{Wald:1993nt,Iyer:1994ys, Barnich:2001jy}. We show that the $T\bar T$ spectrum with $\mu > 0$ can be recovered from the quasilocal energy at a cutoff surface in the auxiliary part of an extended BTZ spacetime. In addition, we derive the redshifted temperatures of the BTZ black hole at the cutoff surface, show that they are compatible with the first law of thermodynamics, and reproduce the bound on the temperatures of $T\bar T$-deformed CFTs.

\subsection{The covariant formulation of gravitational charges}

Let us consider a gravitational theory whose fields are collectively denoted by $\phi$. Given a solution to the equations of motion $\phi$, we denote a nearby point in the phase space of solutions by $\phi + \delta \phi$. The infinitesimal variation of the Hamiltonian $\mathscr H^{\Sigma}_\xi[\phi]$ generating a diffeomorphism $x^\mu \to x^\mu +\xi^\mu$ in a spatial region $\Sigma$ is then given by \cite{Wald:1993nt,Iyer:1994ys, Barnich:2001jy}
\eq{
\delta \mathscr H^{\Sigma}_\xi[\phi]&=\int_\Sigma \bm\omega[\phi,\delta \phi,\delta_\xi \phi],\label{charge}
}
where $\bm\omega[\phi,\delta \phi,\delta_\xi \phi]$ is the symplectic form of the gravitational theory and $\delta_\xi \phi$ denotes the infinitesimal action of the diffeomorphism on $\phi$. The symplectic form is closed on-shell, which in the absence of topological obstructions can be written as $\bm\omega[\phi,\delta \phi,\delta_\xi \phi] =  d \bm \chi_\xi[\phi, \delta \phi]$ where $ \bm \chi_\xi[\phi, \delta \phi]$ is given in pure three-dimensional Einstein gravity by
\eq{
\!\!\! {\bm \chi}_{\xi} [g, \delta g]  \! &= \! \frac{\sqrt{-g}\,\epsilon_{\mu\nu\a}}{16\pi G}  \big[ \xi^\mu (\nabla_\sigma \delta g^{\nu\sigma} \! - \! \nabla^\nu \delta g_\sigma^{\; \sigma}) \! +\xi_\sigma\nabla^\nu \delta g^{\mu\sigma} \! + ( \tfrac{1}{2}  \delta g_\sigma^{\; \sigma} \nabla^\nu  \! - \!  \delta g^{\sigma \nu}\nabla_\sigma)\,\xi^\mu \big] dx^\a\!.
  \label{chidef}
  }
The variation of the Hamiltonian \eqref{charge} can then be written as a surface integral
\eq{
\delta \mathscr H^{\Sigma}_\xi[\phi] & =  \int_{\p\Sigma} \bm \chi_\xi[\phi, \delta \phi] ,\label{charge2}
}
where $\partial\Sigma$ denotes the boundary of the co-dimension one spatial surface $\Sigma$.

When $\xi$ is a Killing vector, the symplectic form vanishes, and the variation of the Hamiltonian is exactly zero. In this case, the infinitesimal variation of the gravitational charge $\mathcal Q_\xi^{\mathcal C}[\phi]$ associated with the symmetry generated by $\xi$ can be defined as
\eq{
 \d \mathcal Q_{\xi}^{\,\mathcal C} [\phi] = \int_{\mathcal C} \bm \chi_\xi[\phi, \delta \phi], \label{surfacecharge}
}
where $\mathcal C$ is a co-dimension two surface. By Stoke's theorem, the infinitesimal charge \eqref{surfacecharge} is independent of the choice of surface $\mathcal C$ provided that any other choice is homologous to the original one. In particular, when $\mathcal C$ is a closed surface at the asymptotic boundary, the gravitational charge \eqref{surfacecharge} yields the ADM mass when $\xi = \p_t$ or the angular momentum when $\xi = \p_\vp$.  For stationary solutions, these asymptotic charges can be calculated on any surface $\mathcal C$ as long as the surface is homologous to $\mathcal C_\infty$.


\subsection{Quasilocal energy and angular momentum}

An important piece of evidence for the correspondence between cutoff AdS and \TTbar-deformed CFTs is the agreement between the \TTbar spectrum \eqref{ttbarspectrum} and the quasilocal energy of BTZ black holes \cite{McGough:2016lol}. The quasilocal energy is the total energy in the region enclosed by the cutoff surface, the latter of which has been computed in the $\mu < 0$ case using the Brown-York stress tensor~\cite{Brown:1992br,Brown:1994gs}. In this section we use the covariant formalism to compute the quasilocal energy and angular momentum of BTZ black holes at a cutoff surface in the $\rho < 0$ and $\rho > 0$ regions of the extended BTZ spacetime \eqref{extendedbtz}. We will show that the energy and angular momentum of these solutions matches the spectrum of \TTbar-deformed CFTs with either sign of the deformation parameter provided that the cutoff radius is identified with the $T\bar T$ deformation parameter via \eqref{generalcutoff}.

Before the deformation, states in the boundary CFT are dual to asymptotically AdS$_3$ spacetimes satisfying Brown-Henneaux boundary conditions \cite{Brown:1986nw}. In the Fefferman-Graham gauge, the most general solution to Einstein's equations satisfying these boundary conditions at $\rho \to 0^+$ is given in \eqref{banados}. For the BTZ black hole, where the $\mathcal L(u)$ and $\bar{\mathcal L}(v)$ functions are constant and parametrized by \eqref{btzparam}, the energy and angular momentum are given by the gravitational charges \eqref{surfacecharge} associated with the $\p_{t}$ and $\p_\vp$ Killing vectors, namely
\eq{
E(0) \equiv \ell^{-1}\mathcal Q_{ \p_{t}}^{\infty} = \frac{c}{6\ell } (T_u^2 + T_v^2), \qquad J(0) \equiv \mathcal Q_{ \p_{\vp}}^{\infty} = \frac{c}{6} (T_u^2 - T_v^2), \label{btzcharges}
}
where $c = 3\ell/2G$ is the central charge of the dual CFT. Note that the gravitational charges are evaluated at asymptotic infinity where $\p_{t}$ and $\p_\vp$ are the generators of time translations and rotations associated with an observer at the asymptotic boundary.

After the deformation, we interpret the effect of the \TTbar deformation as moving the asymptotic boundary of \eqref{extendedbtz} to a cutoff surface in the interior of either the original or auxiliary parts of the spacetime according to \eqref{cutoff-glueon}. The quasilocal energy and angular momentum are defined as the total energy and angular momentum in the region enclosed within the cutoff surface, which are then identified with the energy and angular momentum of the dual \TTbar-deformed CFT. On the other hand, we have seen that in the covariant formalism the gravitational charges associated with an isometry of the spacetime can be evaluated on any surface that is homologous to the cutoff surface. This means, in particular, that we can evaluate the quasilocal energy and angular momentum at the asymptotic boundary. It is natural to wonder how these statements are compatible with each other, namely, how the covariant charges depend on the location $\rho_c$ of the cutoff surface. We will see that the cutoff dependence is encoded in the Killing vectors used to measure the energy and momentum at the cutoff surface.

The holographic dictionary instructs us to identify the locally flat coordinates at the cutoff surface \eqref{cc1} and \eqref{cc2} with the coordinates of the dual field theory. This means that the deformed energy and angular momentum correspond to the energy and angular momentum measured by locally inertial observers at the cutoff surface. Consequently, the variation of the deformed (or quasilocal) energy and angular momentum are given in terms of the gravitational charges by\footnote{The superscript indicates that these charges are evaluated at the cutoff surface. As noted above, these charges depend on the cutoff surface only through the Killing vectors.}
\eq{
\delta E(\mu) \equiv  \ell^{-1} \d \mathcal Q^{\,\rho_c}_{\xi_{t'}} [g], \qquad \delta J(\mu) \equiv  \d \mathcal Q^{\,\rho_c}_{\xi_{\vp'}} [g],
}
where $\xi_{t'}$ and $\xi_{\vp'}$ are the Killing vectors generating time translations and rotations at the cutoff surface. These Killing vectors can be determined from the change of coordinates \eqref{cc1} and \eqref{cc2}, or alternatively, from the following relations
\eq{
\xi_{\vp'}=\p_\varphi, \qquad \xi_{\vp'} \cdot \xi_{t'}=0, \qquad \xi_{t'} \cdot \xi_{t'} = - \frac{1}{\zeta_c},
}
where it is assumed that $\xi_{t'}$ has components only along the $u$ and $v$ coordinates (in the Fefferman-Graham gauge). In addition, we require $\xi_{t'}$ to point in the same direction as $\p_{t}$ such that $\xi_{t'}$ reduces to $\p_{t}$ when the cutoff surface is taken to the asymptotic boundary. Note that the Killing vectors $\xi_{t'}$ and $\xi_{\vp'}$ depend on the choice of radial function since the latter can affect the size of the spatial circle. For the BTZ black hole with the choice \eqref{choicezeta}, the vector generating translations along $t'$ is found to be given by
\eq{
\xi_{t'} =\frac{
	\big(1 + \zeta_c (T_v^2 - T_u^2)\big) \p_u
	- \big(1 + \zeta_c (T_u^2 - T_v^2)\big)\,\p_v
}{\ell \sqrt{ \big(1 - \zeta_c (T_u + T_v)^2 \big) \big(1 - \zeta_c (T_u - T_v)^2 \big) }}.
\label{localstatic}
}

When evaluating the conserved charges, it is important to impose the correct boundary conditions on the variation of the metric at the cutoff surface, namely,
\eq{
\delta (g_{\mu\nu} dx^\mu dx^\nu)\big|_{\rho_c} = 0. \label{cutoffvar}
}
We have written the boundary condition \eqref{cutoffvar} in a manifestly gauge-invariant way. Note that the left-hand side of \eqref{cutoffvar} differs from the naive variation of the metric components $\d g_{\mu\nu}$, the latter of which are not covariant under solution-dependent coordinate transformations and are generically non-zero at the cutoff surface. Relatedly, note that $\d \rho_c \ne 0$ since $\rho_c$ in \eqref{cutoff-glueon} depends on the phase space parameters $T_u$ and $T_v$. In order to circumvent these obstacles we can switch to a new set of coordinates $(u',v',\rho')$ where $\rho'_c$ is independent of $T_u$ and $T_v$ while $(u', v')$ is given by \eqref{cc1} and \eqref{cc2}. In this case, \eqref{cutoffvar} reduces to $\d g'_{\mu'\nu'}|_{\rho'_c} = 0$, which ensures Dirichlet boundary conditions at the cutoff surface. This is the method used, for example, in \cite{Kraus:2021cwf}. Alternatively, if we do not change to the primed coordinates, we can include the variation of the original coordinates $(\var{u},\var{v},\var{\rho})$ in the variation of the metric following \eqref{cutoffvar}. Effectively, this adds improvement terms to $\d g_{\mu\nu}$ so that it transforms covariantly under $T_{u,v}$-dependent coordinate transformations. Both of these approaches are equivalent and produce the same infinitesimal charge $\var{\mcal{Q}_{\xi}^{\,\mathcal C}}$.

The infinitesimal variations of the deformed energy and angular momentum associated with the $\xi_{t'}$ and $\xi_{\vp'}$ Killing vectors are found to be given by
\eqsp{
\delta E(\mu) & = -\frac{c}{6} \frac{1}{\zeta_c} \, \delta \Big( \sqrt{ \big(1 - \zeta_c (T_u + T_v)^2 \big) \big(1 - \zeta_c (T_u - T_v)^2 \big) }\, \Big),
\\[.5ex]
\d J(\mu) &= \frac{c}{6} \, \delta  (T_u^2 - T_v^2 ),   \label{btzchargesdef}
}
where the variation is restricted to the space of solutions parametrized by $T_u$ and $T_v$, namely $\delta = \var{T_u}  \pd_{T_u} + \var{T_v} \pd_{T_v}$. The resulting $\var{E}(\mu)$ and $\var{J}(\mu)$ are manifestly integrable in the space of solutions, which provides a basic consistency check of our approach. In particular, we note that the constraint \eqref{cond1}, which is necessary for a real value of the cutoff radius, guarantees that the spectrum is real when $\mu < 0$ ($\zeta_c > 0)$. The deformed energy and angular momentum are determined from \eqref{btzchargesdef} up to integration constants. The latter are fixed by requiring the deformed charges to reduce to the undeformed ones \eqref{btzcharges} in the limit $|\mu| \to 0$ ($|\zeta_c| \to 0$). In terms of the undeformed charges \eqref{btzcharges}, the deformed energy and angular momentum of the BTZ black hole can be written as
\eq{
E(\mu)= -\frac{\ell}{2\mu} \bigg( 1-\sqrt{1 + \frac{4 \mu}{\ell} E(0) + \frac{4\mu^2}{\ell^4} J(0)^2 }\,\bigg), \qquad J(\mu) = J(0), \label{deformedcharges}
}
where we used the holographic dictionary \eqref{dictionary}.

The deformed energy and angular momentum \eqref{deformedcharges} are valid for both signs of the deformation parameter and agree with the spectrum of \TTbar-deformed CFTs given in \eqref{ttbarspectrum}. The advantage of using the covariant formalism is that it can be employed in more general setups where the analog of the Brown-York stress tensor is not known, e.g.~in the single-trace version of the \TTbar deformation~\cite{Apolo:2019zai}.


\subsection{Thermodynamics}

Let us now show that the quasilocal energy and angular momentum \eqref{deformedcharges} are consistent with the thermodynamics of the BTZ black hole at the cutoff surface. In addition, we will show that the deformed temperatures measured by observers at the cutoff surface reproduce the bound on the temperatures found in the dual field theory.

The BTZ black hole features a horizon at $\rho_h = 1/T_u T_v$ with a thermal identification of the lightcone coordinates
\eq{
(u,v) \sim (u -  i/\ell T_L(0), \,v  +  i/ \ell T_R(0)),  \label{TLTR0}
}
where $T_L(0) = T_u/\ell \pi$ and $T_R(0) = T_v/ \ell \pi$ are interpreted as left and right-moving temperatures in the dual CFT at the asymptotic boundary. After the deformation, we can imagine the $T\bar T$-deformed CFT as living at a cutoff surface in the $\rho < 0$ or $\rho > 0$ regions of the extended black hole geometry. This leads to a redshift of the asymptotic temperatures measured by locally inertial observers at the cutoff surface such that
\eq{
(u', v') \sim (u' -  i/\ell T_L(\mu), \,v' + i / \ell T_R(\mu)).
}
Using the expressions for the locally flat coordinates \eqref{cc1} and \eqref{cc2} we find that the deformed temperatures $T_{L,R}(\mu)$ are given by
\eqsp{
\frac{1}{T_L(\mu)} + \frac{1}{T_R(\mu)} &=
\ell \pi\,\Big( \frac{1}{T_u} + \frac{1}{T_v} \Big)
\sqrt{ \big(1 - \zeta_c (T_u+T_v)^2 \big) \big(1 - \zeta_c (T_u-T_v)^2 \big) }, \\
\frac{1}{T_L(\mu)} - \frac{1}{T_R(\mu)} &=
\ell \pi\,\Big( \frac{1}{T_u} - \frac{1}{T_v} \Big)
\big(1 - \zeta_c (T_u+T_v)^2 \big). \label{Tbulk}
}
These can be inverted so that $T_{u,v}$ read
\eq{
	T_u + T_v = \pi\ell\,\frac{T_L(\mu) + T_R(\mu)}{\gamma(\mu)},
\qquad
	T_u - T_v = \pi\ell\,\big(T_L(\mu) - T_R(\mu)\big),
\label{Tbulk2}
}
where $\gamma(\mu)$ is given by
\eq{
\gamma(\mu) = \sqrt{\vphantom{\big)}1 + 4\pi^2 \ell^2 \zeta_cT_L(\mu) T_R(\mu)}.
}
The entropy of the BTZ black hole can be written in terms of the undeformed and deformed temperatures as
\eq{
S = \frac{ \pi^2\ell  c}{3}[T_L(0) + T_R(0)] = \frac{\pi^2\ell  c}{3} \bigg[ \frac{T_L(\mu) + T_R(\mu)}{\gamma(\mu)}\bigg], \label{entropy}
}

Although the numerical value of the entropy has not changed, its dependence on the deformed temperatures has been modified. The result \eqref{entropy} matches the expression for the entropy in $T\bar T$-deformed CFTs derived in \cite{Apolo:2019zai}. In addition, using the infinitesimal charges \eqref{btzchargesdef},  it is not difficult to check that the temperatures \eqref{Tbulk} are consistent with first law of thermodynamics, namely
\eq{
\delta S = \frac{1}{T_L(\mu)} \delta E_L(\mu)+ \frac {1}{T_R(\mu)}\delta E_R(\mu).
}
where $E_{L,R} = \frac{1}{2}(E \pm J)$ denote the left and right-moving energies conjugate to the left and right-moving temperatures $T_{L,R}$.

It is also interesting to calculate the product of the deformed temperatures which reads
\eq{
T_L(\mu)\, T_R(\mu)= \frac{1}{\pi^2 \ell^2} \frac{T_uT_v }{[1 - \zeta_c (T_u + T_v)^2]}. \label{TLTRprod}
}
When $\mu < 0$ ($\zeta_c > 0$), this expression is unbounded from above and tends to infinity as $\zeta_c \to 1/(T_u + T_v)^2$. This limit corresponds to a cutoff surface that approaches the outer horizon of the BTZ black hole. On the other hand, using the holographic dictionary \eqref{dictionary}, we find that when $\mu > 0$ ($\zeta_c < 0$) the temperatures \eqref{Tbulk} satisfy
\eq{
T_L(\mu)\,T_R(\mu) \le - \frac{1}{4\pi^2 \ell^2 \zeta_c} = \frac{3}{4\pi^2c\mu} = T_H(\mu)^2,
}
which is precisely the field theory relation \eqref{productbound}, thus providing a geometric derivation of the Hagedorn temperature \eqref{productbound}. In particular, we see from \eqref{entropy} that exceeding the Hagedorn temperature corresponds to a complex entropy.


\section{\TTbar partition functions from cutoff/glue-on AdS} \label{se:partitionfunction}

In this section we consider the partition function of $T\bar T$-deformed CFTs from the bulk and boundary sides of the correspondence. We first review the field theory derivation of the torus partition function and provide a derivation of the sphere partition function, results that are valid at large $c$. We then show that the contributions of thermal AdS and Euclidean BTZ to the gravitational path integral match the torus partition function of $T\bar T$-deformed CFTs. Furthermore, we compute the contribution of the sphere foliation of Euclidean AdS to the path integral, and show that it reproduces the regularized partition function on the sphere.


\subsection{Field theory derivations} \label{se:TTbarpartitionfunctions}

In this section we consider the field theory derivations of the torus and sphere partition functions of $T\bar T$-deformed CFTs. The definition of the \TTbar deformation \eqref{TTbardef} implies that the partition function must satisfy the following differential equation
\eq{
\partial_\mu \log Z_{\TTbar}(\mu)&=8\pi \int d^2x\sqrt{\gamma}\, \langle T\bar{T} \rangle. \label{diff1}
}
This equation differs from \eqref{TTbardefintro} by a minus sign due to the Euclidean signature. In the following, we first review the torus partition function derived in \cite{Apolo:2023aho}, and then derive the sphere partition function by solving a set of differential equations that includes \eqref{diff1}. Both of these derivations are valid in the semiclassical limit where the central charge of the undeformed CFT is large. In particular, we comment on the role of initial conditions in determining the sphere partition function, and interpret previous results found in the literature.

\subsubsection*{The $T \bar T$ partition function on the torus}

Let us consider a $T\bar T$-deformed CFT on a torus with inverse left and right-moving temperatures $\beta_{L,R}$. At low temperatures $\beta_{L,R} \gg 1$, the torus partition function is dominated by the contribution from the vacuum state. At high temperatures $\beta_{L,R} \ll 1$, modular invariance implies that the partition function is dominated by the modular image of the vacuum~\cite{Datta:2018thy}. In analogy with two-dimensional CFTs \cite{Hartman:2014oaa}, the torus partition function of $T\bar T$-deformed CFTs can also be approximated in the semiclassical regime where the central charge of the undeformed CFT is large. In this case, modular invariance and a sparse spectrum of light states implies that the torus partition function is universally given by \cite{Apolo:2023aho}
\eq{
\log   Z_{T\bar T}(\mu)  \approx \left\{ \begin{aligned}
& {-\frac{1}{2}}\,(\beta_L + \beta_R)\, RE_{\text{vac}}(\mu),  &\beta_L \beta_R > 1, \\
& {-2 \pi^2 \bigg(\frac{1}{\beta_L} + \frac{1}{\beta_R}\bigg)  RE_{\text{vac}}\bigg(\frac{4\pi^2}{\beta_L \beta_R} \mu \bigg)}, \ \quad &\beta_L\beta_R < 1,
 \end{aligned} \right. \label{ZTTbar}
}
where $E_{\text{vac}}(\mu)$ is the energy of the vacuum given in \eqref{Evac}. According to the holographic dictionary, the bulk description in terms of semiclassical Einstein gravity is valid when the central charge of the undeformed CFT is large, namely when $c = 3\ell/2G \gg 1$. Hence, \eqref{ZTTbar} is the appropriate expression for comparison with the gravitational path integral.


\subsubsection*{The $T \bar T$ partition function on the sphere}

Let us now consider a $T\bar T$-deformed CFT on the two-dimensional sphere of radius $L$,
\eq{
ds^2 = L^2(d\theta^2+\sin^2\theta d\phi^2).
}
The sphere partition function must satisfy the following partial differential equation, in addition to the $T\bar T$ differential equation \eqref{diff1}
\eq{
-L\,\partial_L \log Z_{\TTbar}(\mu)&=\int d^2x\sqrt{\gamma}\,\langle T^i_i \rangle. \label{diff2}
}
A crucial property of $T\bar T$-deformed CFTs is the factorization of the expectation value of the $T\bar T$ operator. With some exceptions \cite{Brennan:2020dkw}, the $T\bar T$ operator does not factorize in general curved spacetimes \cite{Jiang:2019tcq}. Nevertheless, if one appeals to large-$c$ factorization, which is a necessary condition for the undeformed CFT to be holographic, then the factorization of $\langle T\bar T\rangle$ can be assumed to hold. If we also assume rotational symmetry along $\phi$, then the conservation law of the stress tensor and the trace flow equation \eqref{floweq} can be shown to be solved by \cite{Donnelly:2018bef}
\eq{\label{sphere:stresstensor}
\langle T_{ij}\rangle =-\frac{1}{4\pi\mu} \bigg(1-\sqrt{1-\frac{c\mu}{3L^2}}\, \bigg)  \gamma_{ij}.
}
In analogy with \eqref{criticalmu}, the stress tensor \eqref{sphere:stresstensor} indicates an upper bound of the deformation parameter on a sphere with radius $L$, namely
\eq{
\mu < \mu_c \equiv \frac{3L^2}{c}.\label{mucsphere}
}
Using the expectation value of the stress tensor \eqref{sphere:stresstensor}, the general solution to the differential equations \eqref{diff1} and \eqref{diff2} is given by
\eq{
\log Z_{\TTbar}(\mu, a) = \frac{c}{3} \log \bigg[\frac{L}{a}   \bigg(1+\sqrt{1-\frac{c \mu }{3  L^2}}\, \bigg) \bigg] - \frac{L^2}{\mu}  \sqrt{1-\frac{c \mu }{3 L^2}} + \frac{L^2}{\mu}, \label{Zsol}
}
where $a$ is an arbitrary integration constant with the dimension of length.

The  choice of integration constant in \eqref{Zsol} depends on the initial conditions of the flow. For example, it is natural to require that the \TTbar-deformed partition function reduces to the partition function of the undeformed CFT in the limit $\mu \to 0$. In this limit, the partition function \eqref{Zsol} becomes
\eq{
\lim_{\mu \to 0}\log Z_{T\bar T}(\mu ,a) =\frac{c}{6} \pqty{1 + \log \frac{4L^2}{a^2}}. \label{Z0}
}
It is then natural to identify $a$ with the UV cutoff $\epsilon$, case in which the partition functions of the deformed and undeformed theories are UV divergent. Alternatively, we can consider a renormalized partition function such that $a$ is finite and its value depends on the renormalization scheme. We note that the two choices of the integration constant are compatible with the sphere partition function of Liouville CFT at large $c$, before and after regularization, which was explicitly computed in \cite{Zamolodchikov:2007abc}.

The sphere partition function of $T\bar T$-deformed CFTs with $\mu < 0$ has been previously computed in \cite{Donnelly:2018bef,Li:2020zjb}. However, we note that the partition function of \cite{Donnelly:2018bef} does not satisfy the $T\bar T$ differential equation \eqref{diff1}. This can be understood as a consequence of identifying the energy scale of the theory with the deformation parameter, which is motivated by the cutoff AdS proposal. Under this assumption, the UV cutoff changes along with $\mu$, reason why the partition function satisfies a modified differential equation
\eq{
2\mu\,\partial_\mu \log Z_{\TTbar}(\mu)=
-L\,\partial_L \log Z_{\TTbar}(\mu ) =\int d^2x\sqrt{\gamma}\,\langle T^i_i \rangle.\label{floweqZ}
}
Integrating the above equations, together with the initial condition $\log Z_{\TTbar}(\mu) = 0$ at $L=0$, it is not difficult to show that the resulting partition function can still be written as \eqref{Zsol}, but with the integration constant given by $a = \sqrt{c|\mu|/3}$.


\subsection{The gravitational on-shell action}

In this section we provide a prescription for the calculation of the Euclidean on-shell action of pure Einstein gravity in glue-on AdS$_3$ spacetimes.

Let us consider the slicing of three-dimensional metrics by a one-parameter family of hypersurfaces $\mathcal N_\zeta$ labelled by $\zeta$ as in \eqref{glueonAdS}, which we reproduce here for convenience
\eq{
ds^2= h_{ij}dx^i dx^j+n_\mu n_\nu dx^\mu dx^\nu, \qquad h_{ij}= \frac{1}{\zeta} \gamma_{ij}, \qquad \zeta \in \mathbb R, \label{metric:AdS}
}
where $h_{ij}$ is the induced metric of $\mathcal N_\zeta$, $n^\mu$ is the unit vector normal to $\mathcal N_\zeta$, and the asymptotic boundary of AdS$_3$ (AdS$_3^*$) is denoted by $\mathcal N_\epsilon$ ($\mathcal N_{-\epsilon}$) and is located at $\zeta = \epsilon$ ($\zeta = -\epsilon$) with $\epsilon \to 0$. In addition, it is convenient to introduce the following notation for the different components of the action of pure three-dimensional gravity
\eq{
I_{\mcal{M}}(\zeta_1, \zeta_2) & \coloneqq - \frac{1}{16\pi G} \int_{\zeta_1}^{\zeta_2} d\zeta \int d^2 x \sqrt{g} \, (R + 2 \ell^{-2}),  \label{bulkaction} \\
I_{\mathcal N_{\zeta}} & \coloneqq  - \frac{1}{8\pi G} \int_{\mathcal N_{\zeta}} d^2x \sqrt{h}\,h^{ij} K_{ij} + \frac{\sigma_\zeta}{8\pi G} \int_{\mathcal N_{\zeta}} d^2x \sqrt{h} \, \bigg(\ell^{-1} - \frac{\ell  \mathcal{R}[h]}{4} \log |\zeta| \bigg), \label{boundaryaction}
}
where $\mcal{M}$ is a three-dimensional manifold, while $K_{ij}$ and $\mathcal{R}[h]$ are the extrinsic curvature and the Ricci scalar of the surface $\mathcal N_{\zeta}$ computed with respect to the induced metric $h_{ij}$. The bulk term \eqref{bulkaction} is the Einstein-Hilbert action with a negative cosmological constant evaluated on $\mcal{M}$, with the radial coordinate integrated from $\zeta_1$ to $\zeta_2$. Depending on the choice of $\mcal{M}$, the boundary $\p\mcal{M}$ generally contains one or two surfaces $\mathcal N_{\zeta_1}$ and $ \mathcal N_{\zeta_2} $, and we should add the boundary terms \eqref{boundaryaction} accordingly. The first term in the boundary action \eqref{boundaryaction} is the Gibbons-Hawking term that guarantees a well-defined variational principle, while the second term corresponds to the counterterms that regularize the otherwise divergent action at $\zeta \to 0^\pm$ \cite{Henningson:1998gx,deHaro:2000vlm}. Note that the $\log |\zeta|$ counterterm, which is only present when the surface $\mathcal N_\zeta$ is curved, is scheme dependent and has been written for a metric of the form \eqref{metric:AdS}. The lack of diff invariance of this counterterm is known to be a reflection of the Weyl anomaly.

Let us now describe how the on-shell action can be evaluated on glue-on Euclidean geometries. When the cutoff surface is located in the AdS$_3$ part of \eqref{metric:AdS}, the action consists of a bulk integral from the cutoff surface $\mathcal N_c$ at $\zeta = \zeta_c$ to the origin of coordinates $\zeta_I > \zeta_c$ such that
\eq{
I(\zeta_c) &= I_{\text{AdS}}(\zeta_c, \zeta_I) + I_{\mathcal N_{c}}, \qquad \zeta_c > 0. \label{actioncutoff}
}
Note that $\zeta = \zeta_I$ is the origin of the radial coordinate and not a boundary, reason why no boundary terms are included there. The action \eqref{actioncutoff} is finite when the cutoff surface $\mathcal N_c$ is pushed to the asymptotic boundary such that $\zeta_c = \epsilon$ with $\epsilon \to 0^+$, where it reduces to the regularized action of pure three-dimensional gravity. In this limit, the bulk action features a logarithmic divergence, which is cancelled by the $\log$ counterterm in \eqref{boundaryaction}, and is a consequence of the Weyl anomaly of the dual CFT.

On the other hand, in the glue-on geometry the action must include a bulk term for each side of the extended space. Furthermore, since the AdS$_3$ region has an asymptotic boundary $\mathcal N_{\epsilon}$, it is natural to include a copy of \eqref{boundaryaction} there. Similarly, since we are imposing a cutoff in the AdS$_3^*$ part of the space, it is natural to include two copies of \eqref{boundaryaction}, one at the asymptotic boundary $\mathcal N_{-\epsilon}$ and one at the cutoff surface $\mathcal N_c$. Therefore, the on-shell action when $\mathcal N_{c} \in \text{AdS}_3^*$ is proposed to be given by
\eq{
\!\! I(\zeta_c) = I_{\text{AdS}^*}(\zeta_c,  -\epsilon) + I_{\mathcal N_{c}} - I_{\mathcal N_{-\epsilon}} + I_{\text{AdS}}(\epsilon, \zeta_I) + I_{\mathcal N_{\epsilon}}, \qquad \zeta_c < 0, \label{actionglueon}
}
where $\epsilon$ is a cutoff that regularizes the divergences of the bulk integrals. The boundary terms $I_{{\mathcal N}_{\pm\epsilon}}$ respectively cancel the divergences from the bulk integrals $I_{\text{AdS}}(\epsilon, \zeta_I)$ and  $ I_{\text{AdS}^*}(\zeta_c,  -\epsilon)$ as $\epsilon\to0$, while the finite contributions from the two asymptotic boundaries cancel each other. The net effect is that the on-shell action only receives contributions from the cutoff surface at $\zeta_c$. This is consistent with the glue-on proposal in section \ref{se:glueonproposal}, and the covariant charge calculation in section \ref{se:charges}, where we have regarded the cutoff surface $\mathcal{N}_{\zeta_c}$ as the boundary of the entire glue-on spacetime. Furthermore, it is straightforward to verify that variation of the actions \eqref{actioncutoff} and \eqref{actionglueon} with respect to $\gamma_{ij}$ correctly reproduces the stress tensor \eqref{BYtensor}.

In analogy with the AdS$_3$/CFT$_2$ correspondence, the partition function of $T\bar T$-deformed CFTs on $\mathcal N_{c}$ is given by the gravitational path integral $\mathcal Z$ over spaces of the form \eqref{metric:AdS} with a finite cutoff such that
\eq{
Z_{T\bar T} (\mu) =  \mathcal Z (\zeta_c). \label{partition}
}
In the semiclassical limit, where the curvature scale is much larger than the Planck length, the right-hand side of \eqref{partition} can be approximated by the on-shell action of the dominant saddle such that
\eq{
Z_{T\bar T} (\mu) =  \mathcal Z (\zeta_c) \approx  e^{-I(\zeta_c)}, \qquad c = \frac{3\ell}{2 G_N} \gg 1,\label{partition2}
}
where the action is given by \eqref{actioncutoff} or \eqref{actionglueon} depending on the location of the cutoff. In what follows, we will calculate on-shell actions on different classical solutions and show that they reproduce the torus and sphere partition functions of $T\bar T$-deformed CFTs in the semiclassical limit where the central charge of the undeformed CFT is large.


\subsection{The torus partition function}

In this section we evaluate the on-shell action on Euclidean AdS$_3$ $\cup$ AdS$_3^*$ spaces with the topology of a torus and a finite cutoff in either AdS$_3$ or AdS$_3^*$. In this case, the gravitational path integral receives contributions from two kinds of saddles: cutoff or glue-on versions of thermal AdS$_3$ and the BTZ black hole.

\subsubsection*{Thermal AdS$_3$}

We begin by describing the extended version of thermal AdS$_3$. For convenience we work in the following gauge\footnote{The relationship between $(t,\phi,r)$ and the $(u,v, \rho)$ coordinates used in \eqref{banados} is given by $r = \rho^{-1/2} - \rho^{1/2}/4$ and $(u, v) = (\varphi + t, \varphi -t)$.}
\eq{
\textrm{AdS}_3: \quad ds^2 = \ell^2 \Big[ (r^2 + 1)\, dt^2 + \frac{dr^2}{r^2 + 1} + r^2 d \vp^2 \Big], \label{globalAdS}
}
where the coordinates satisfy
\eq{
(t, \vp) \sim \bigg(t + \frac{\beta_u + \beta_v}{2}, \, \vp - i \,\frac{\beta_u - \beta_v}{2} \bigg) \sim (t, \vp + 2\pi), \label{torus}
}
with $\beta_{u,v}$ denoting the left and right-moving inverse temperatures. In order to extend \eqref{globalAdS} beyond the asymptotic boundary, we must first make a choice of the radial function $\zeta$ in \eqref{cutoffg} and perform the analytic continuation. Different choices of the radial function correspond to different gauges for the resulting auxiliary spacetime. One convenient choice is to let $\zeta=r^{-2} \in \mathbb{R}$. In this case, the auxiliary AdS$_3^*$ space corresponds to the region with $\zeta<0$, or equivalently $r= i \til r$ where $\til r \in \mathbb R_{\ge0}$. The extended thermal AdS$_3$ space is then given by AdS$_3$ $\cup$ AdS$_3^*$ where AdS$_3^*$ satisfies the same identification of coordinates as \eqref{torus} and reads
\eq{
\textrm{AdS}^*_3: \quad ds^2 = \ell^2 \Bigl[ - (\til r^2 - 1)\, dt^2 + \frac{dr^2}{\til r^2 - 1} - \til r^2 d \vp^2 \Big] . \label{auxglobalAdS}
}
The AdS$_3$ and auxiliary AdS$_3^*$ spaces are glued along their asymptotic boundaries at $r \to \infty$ and $\til r \to \infty$. Note that the $t$ and $\tilde r$ coordinates exchange roles at $\tilde r = 1$. Consequently, the range of the radial coordinate on AdS$_3^*$ is $\tilde{r} \ge 1$. Although both the $t$ and $\vp$ coordinates are timelike in AdS$_3^*$, the metric the $T\bar T$-deformed theory couples to is identified with $ds_c^2$ in \eqref{cutoffmetric}, where both of these coordinates are spacelike.

Using the holographic dictionary \eqref{dictionary}, we see that for $\mu < 0$ the cutoff surface $\mathcal N_{c}$ lies in the AdS$_3$ region at $r^2 = \zeta_c^{-1} = -3 \ell^2/c \mu$. On the other hand, for $\mu > 0$ the cutoff surface $\mathcal N_{c}$ is located in the auxiliary part of the space at $\til r^2 = -\zeta_c^{-1} =3 \ell^2/c \mu $. In particular, the bound $\tilde{r} \ge 1$ implies a minimum value for the cutoff radius such that $\zeta_c \ge -1$, which reproduces the bound on the deformation parameter \eqref{criticalmu}. We see that the bound \eqref{criticalmu} is obtained both in the field theory and bulk sides of the correspondence by studying the consistency of the vacuum.

The on-shell action when the cutoff surface $\mathcal N_c$ lies in either the AdS$_3$ or AdS$_3^*$ regions of the space can be evaluated using \eqref{actioncutoff} or \eqref{actionglueon}, respectively, and is given by
\eq{
I_{\text{AdS}_3}(\zeta_c) &= \frac{c\,(\beta_u + \beta_v)}{12} \frac{1}{\zeta_c} \sqrt{1 + \zeta_c}\,\big( 1 - \sqrt{1 + \zeta_c} \,\big).
}
Then, using the relation between $\zeta_c$ and $\mu$ \eqref{dictionary}, we find that the on-shell action of thermal cutoff/glue-on AdS$_3$ reads
\eq{
I_{\text{AdS}_3}(\mu) =  - \frac{c\,(\beta_u + \beta_v)}{12}  - \frac{\ell^2 (\beta_u + \beta_v)}{4\mu} \bigg( \sqrt{1 - \frac{c \mu}{3\ell^2}} - 1 \bigg). \label{thermal:logZ}
}
As a consistent check, this expression reduces to the expected answer in thermal AdS$_3$ in the limit $\mu \to 0$, namely $I_{\text{AdS}_3}(0)=-c\,(\beta_u + \beta_v)/24$.

In order to relate \eqref{thermal:logZ} to the partition function of a $T\bar T$-deformed CFT \eqref{ZTTbar}, we must first obtain the relationship between $\beta_{u,v}$ and the physical inverse temperatures $\beta_{L,R}$ of the deformed theory. The reason for the discrepancy is that the thermal circle in \eqref{torus} is measured by an observer at the asymptotic boundary while the inverse temperatures $\beta_{L,R}$ of the $T\bar T$-deformed CFT are measured with respect to an inertial observer at the cutoff surface with metric $ds^2_c = \ell^2 (dt'^2 + d\vp'^2)$. The relationship between the $(t,\vp)$ and $(t',\vp')$ coordinates is given by
\eq{
(t', \vp')  &= \big(\sqrt{1 + \zeta_c}\, t, \,\vp\big). \label{thermal:cc}
}
Consequently, the temperatures are related by
\eq{
\beta_u + \beta_v = \frac{\beta_L + \beta_R}{\sqrt{1 + \zeta_c}}, \qquad \beta_u - \beta_v = \beta_L - \beta_R,
\label{thermal:beta}
}
where $\beta_{L,R}$ are held fixed under the deformation and are taken to be independent of $\mu$. Using the inverse temperatures \eqref{thermal:beta}, the on-shell action of cutoff/glue-on AdS$_3$ can be written as
\eq{
I_{\text{AdS}_3}(\mu) = {\frac{1}{2}\,( \beta_L + \beta_R )\,\ell E_{\text{vac}}(\mu)}, \label{thermal:logZ2}
}
where $E_{\text{vac}}(\mu)$ is the vacuum energy of the $T\bar T$-deformed CFT given in \eqref{Evac}. Eq.~\eqref{thermal:logZ2} is the expected contribution of the vacuum to the partition function of $T\bar T$-deformed CFTs on the torus. In addition, we note that this expression matches the field theory result \eqref{ZTTbar} for small temperatures $\beta_L \beta_R>1$, in the limit where the central charge of the undeformed CFT is large \cite{Apolo:2023aho}.

Finally, we find that the on-shell action on thermal AdS$_3$ satisfies the Euclidean version of the $T\bar T$ differential equation~\eqref{diff1} where the stress tensor in the $(t',\varphi')$ coordinates can be obtained from the Brown-York stress tensor \eqref{BYtensor} by the change of coordinates \eqref{thermal:cc}, namely
\renewcommand{\arraystretch}{1.2}
\eq{
\ave{T_{ij}} &= - \frac{\ell^2}{4\pi\mu}  \left( \begin{array}{cc}
1 - \sqrt{1 - \frac{c\mu}{3\ell^2}} & 0 \\
0 & 1 - \frac{1}{\sqrt{1 - \frac{c\mu}{3\ell^2}}}
\end{array} \right) = \frac{\ell}{2\pi}  \left( \begin{array}{cc}
E_{\text{vac}}(\mu)  \vphantom{\sqrt{1 - \frac{c\mu}{3\ell^2}} }& 0 \\
0 & \ell\p_\ell E_{\text{vac}}(\mu)  \vphantom{ \frac{1}{\sqrt{1 - \frac{c\mu}{3\ell^2}}}}
\end{array} \right). \label{thermal:T}
}
We have written the right hand side of \eqref{thermal:T} in terms of the energy of the vacuum \eqref{Evac}, which matches the expectation value of the (dimensionless) stress tensor in $T\bar T$-deformed CFTs in Euclidean signature \cite{Zamolodchikov:2004ce}. Furthermore, it is not difficult to verify that \eqref{thermal:T} satisfies the trace flow equation \eqref{floweq}, which provides an additional consistency check of our calculations.


\subsubsection*{Euclidean BTZ}

Let us now consider the contribution of the nonrotating BTZ black hole to the gravitational path integral. The latter can be obtained from the on-shell action on the Euclidean version of the extended BTZ space \eqref{extendedbtz} with $T_u = T_v = \pi/ \beta_t$ and the identification of coordinates of a rectangular torus, namely $(u, v) \sim (u - i\beta_t, v + i\beta_t) \sim (u + 2\pi, v + 2\pi)$. Alternatively, we can express the BTZ $\cup$ BTZ$^*$ space in the following gauge
\eq{
\text{BTZ}:& \qquad ds^2 = \ell^2 \bigg[ \frac{d r^2}{r^2 - r_+^2} + (r^2 -  r_+^2) \,dt^2 + r^2 d\vp^2 \bigg], \label{BTZ} \\
\text{BTZ}^*: & \qquad ds^2 = \ell^2 \bigg[ \frac{d \til r^2}{\til r^2 +  r_+^2} - ( \til r^2 +  r_+^2) \,dt^2 - \til r^2 d\vp^2 \bigg], \label{BTZstar}
}
where $(-i t, \vp) = \frac{1}{2} (u - v, u + v)$ and $r_+ = 2\pi/ \beta_t$. The auxiliary BTZ$^*$ space is obtained from BTZ by the same analytic continuation used in \eqref{auxglobalAdS}, namely $r = i \til r$ with $\til r \in \mathbb R_{\ge 0}$. In this gauge, the cutoff surface is located at a constant value of the radial coordinate in either the BTZ ($r^2 = \zeta_c^{-1}$) or BTZ$^*$ ($\til r^2 = -\zeta_c^{-1}$) regions of the background.

The on-shell action of cutoff/glue-on BTZ is given by
\eq{
I_{\text{BTZ}} (\zeta_c) &=  \frac{c \beta_t}{6} \frac{1}{\zeta_c} \bigg(\sqrt{1 - \frac{4\pi^2 \zeta_c}{{\beta_t^2}}} -  1 \bigg). \label{btz:action}
}
In analogy with the discussion in thermal AdS$_3$, we need to express \eqref{btz:action} in terms of the deformed inverse temperature $\beta$ of the $T\bar T$-deformed CFT. The latter can be obtained from \eqref{Tbulk2} such that
\eq{
\beta_t = \sqrt{\beta^2 + 4\pi^2 \zeta_c}.
}
In terms of the deformed temperature and the deformation parameter, the on-shell action of cutoff/glue-on BTZ \eqref{btz:action} can be written as
\eq{
I_{\text{BTZ}}(\mu) = { \frac{4 \pi^2}{ \beta}\ell E_{\text{vac}}\bigg(\frac{4\pi^2}{\beta^2} \mu \bigg)}. \label{btz:Zexp}
}
It is not difficult to verify that \eqref{btz:Zexp} satisfies the $T\bar T$ differential equation \eqref{diff1}, where the stress tensor of the dual $T\bar T$-deformed CFT is given in Cartesian coordinates $(t',\varphi')$ by
\renewcommand{\arraystretch}{1.2}
\eq{
\! \ave{T_{ij}} &= -\frac{\ell^2}{4\pi\mu} \left( \begin{array}{cc}
1 - \frac{1}{\sqrt{1 - \frac{4 \pi^2 c\mu}{3\ell^2\beta^2}}} & 0 \\
0 & 1 - \sqrt{1 - \frac{4 \pi^2 c\mu}{3\ell^2\beta^2}}
\end{array} \right)  = \frac{\ell}{2\pi}  \left( \begin{array}{cc}
E_{\text{BTZ}}(\mu)  \vphantom{\frac{1}{\sqrt{1 - \frac{4 \pi^2 c\mu}{3\ell^2\beta^2}}}}& 0 \\
0 & \ell\p_\ell E_{\text{BTZ}}(\mu)  \vphantom{ \sqrt{1 - \frac{4 \pi^2 c\mu}{3\ell^2\beta^2}}}
\end{array} \right) , \label{btz:T}
}
where $E_{\text{BTZ}}(\mu)$ denotes the deformed energy of the nonrotating BTZ black hole obtained from \eqref{deformedcharges} with $E(0) = \frac{c\,\pi^2}{3\ell\beta_t^2}$ and $J(0) = 0$.

Following the steps described above it is straightforward to show that the on-shell action for both the cutoff/glue-on version of the rotating BTZ black hole is given by
\eq{
I_{\text{BTZ}}(\mu)  = {2 \pi^2 \bigg(\frac{1}{\beta_L} + \frac{1}{\beta_R}\bigg)\,\ell E_{\text{vac}}\bigg(\frac{4\pi^2}{\beta_L \beta_R} \mu \bigg)}, \label{btz:ZexpLR}
}
where $\beta_{L,R}$ denote the deformed inverse temperatures of the dual $T\bar T$-deformed CFT, the latter of which are related to the inverse temperatures $\beta_{u,v} = \pi/T_{u,v}$ of the BTZ black hole by \eqref{Tbulk2}. This expression matches the field theory result \eqref{ZTTbar} in the semiclassical limit when the temperatures are large ($\beta_L \beta_R < 1$). Note that in analogy with the AdS/CFT correspondence, the contributions to the gravitational path integral of the glue-on thermal AdS$_3$ \eqref{thermal:logZ2} and BTZ \eqref{btz:ZexpLR} geometries are related by the generalized modular S transformation of $T\bar T$-deformed CFTs \cite{Datta:2018thy}, namely
\eq{
(\beta_L, \beta_R, \mu) \mapsto  \bigg(\frac{4\pi^2}{\beta_L}, \frac{4\pi^2}{\beta_R}, \frac{4\pi^2 \mu}{\beta_L \beta_R} \bigg).
}

To summarize, using the saddle point approximation we have shown that in the semiclassical limit, the gravitational path integral on extended AdS spaces with a finite cutoff and the topology of the torus is given by
\eq{
\log \mathcal Z(\mu)  \approx - I(\zeta_c) =  \left\{ \begin{aligned}
& {- \frac{1}{2}}\,(\beta_L + \beta_R)\, \ell E_{\text{vac}}(\mu),  &\beta_L \beta_R > 1, \\
& {-2 \pi^2 \bigg(\frac{1}{\beta_L} + \frac{1}{\beta_R}\bigg)  \ell E_{\text{vac}}\bigg(\frac{4\pi^2}{\beta_L \beta_R} \mu \bigg)}, \ \quad &\beta_L\beta_R < 1.
 \end{aligned} \right.  \label{summaryZ}
}
The result \eqref{summaryZ} matches the field theory calculation \eqref{ZTTbar}. In addition, each of the lines in \eqref{summaryZ} satisfies the differential equation \eqref{diff1} with respect to the appropriate stress energy tensor, namely \eqref{thermal:T} or the rotating generalization of \eqref{btz:T}. As a result, we can verify that \eqref{summaryZ} satisfies the defining equation for the $T\bar T$ deformation.


\subsection{The sphere partition function}

In this section we evaluate the on-shell action of the cutoff and glue-on AdS$_3$ spaces with the topology of a sphere at the cutoff surface. Following the general prescription of section \ref{se:glueondef}, the sphere foliation of the glue-on space can be written as
\eq{
\quad ds^2 =  \frac{\ell^2  d \zeta^2}{4\zeta^2(1 + \frac{\ell^2}{L^2}\zeta)} + \frac{L^2}{\zeta} \big( d\theta^2 + \sin\theta^2\,d\phi^2\big), \qquad \zeta \ge \zeta_c=- \frac{c \mu}{3 \ell^2}, \label{adsSphere}
}
where $\theta \in [0, \pi)$, $\vp \sim \vp + 2\pi$, and $L$ is the radius of the sphere where the \TTbar-deformed theory is defined
\eq{
ds_c^2 = \gamma_{ij} dx^i dx^j = L^2 \big( d\theta^2 + \sin\theta^2\,d\phi^2 \big). \label{boundarymetricsphere}
}
Note that the three-dimensional metric \eqref{adsSphere} has a divergent determinant and hence becomes noninvertible at $\zeta = -L^2/\ell^2$. In analogy with thermal AdS$_3$, it is reasonable to prevent the cutoff surface from reaching this value of the radial coordinate such that $\zeta_c=- \frac{c \mu}{3 \ell^2}> -L^2/\ell^2$, which reproduces the bound on the deformation parameter \eqref{mucsphere} derived from the field theory side of the correspondence.

The on-shell action for the sphere foliation of cutoff/glue-on AdS$_3$ reads
\eq{
I_{S^2}(\zeta_c) = - \frac{c}{3} \log \bigg(\frac{L}{\ell}  + \frac{L}{\ell}\sqrt{1 - \frac{c\mu}{3L^2}} \bigg) + \frac{L^2}{\mu} \sqrt{1 - \frac{c\mu}{3L^2}} - \frac{L^2}{\mu},  \label{sphere:logZ}
}
which matches the expression obtained in the field theory side \eqref{Zsol} provided that the integration constant is identified with the scale of AdS, namely $a = \ell$. Note that the $\log$ counterterm in \eqref{boundaryaction}, which is a consequence of the Weyl anomaly, makes a crucial contribution to the on-shell action, as noted previously in \cite{Caputa:2020lpa,Li:2020zjb}. In particular, it guarantees that the partition function is compatible with the $T\bar T$ differential equation \eqref{diff1}, where the stress tensor obtained from \eqref{BYtensor} is given by
\eq{
\ave{T_{ij}} &= -\frac{1}{4 \pi \mu} \bigg( 1 -   \sqrt{1 - \frac{c\mu}{3L^2}} \, \bigg) \gamma_{ij}. \label{sphere:T}
}
Note that the stress tensor matches the field theory derivation in \eqref{sphere:stresstensor}.

As discussed in section \ref{se:TTbarpartitionfunctions}, the sphere partition function of $T\bar T$-deformed CFTs with $\mu < 0$  computed in \cite{Donnelly:2018bef} does not satisfy the $T\bar T$-differential equation as it assumes that the UV cutoff is related to the deformation parameter in a particular way. Nevertheless, we note that this partition function can be obtained from the bulk on-shell action by omitting the $\log$ counterterm associated with the Weyl anomaly \cite{Caputa:2019pam}. This is compatible with the general approach of \cite{Hartman:2018tkw}, which assumes the two aforementioned scales are related to each other. In contrast, our result \eqref{sphere:logZ} is compatible with the general approach used in \cite{Caputa:2020lpa} where the UV cutoff and the deformation parameter are assumed to be independent.	


\bigskip

\section*{Acknowledgments}
We are grateful to Bin Chen, Jin Chen, Zhengyuan Du, Xia Gu, Monica Guica, Kangning Liu, Reiko Liu, Dominik Neuenfeld, Andrew Rolph, Mauricio Romo, Jie-Qiang Wu, Xianjin Xie, Boyang Yu and Yuan Zhong for helpful discussions. LA thanks the Asia Pacific Center for Theoretical Physics (APCTP) for hospitality during the focus program ``Integrability, Duality and Related Topics", as well as the Korea Institute for Advanced Study (KIAS) for hospitality during the ``East Asia Joint Workshop on Fields and Strings 2022'', where part of this work was completed. The work of LA was supported by the Dutch Research Council (NWO) through the Scanning New Horizons programme (16SNH02). The work of PXH, WXL, and WS is supported by the national key research and development program of China No.~2020YFA0713000.



\bigskip

\bibliographystyle{JHEP.bst}
\bibliography{glueon.bib}

\end{document}